\documentclass[fleqn,usenatbib]{mnras}

\usepackage{newtxtext,newtxmath}

\usepackage[T1]{fontenc}

\DeclareRobustCommand{\VAN}[3]{#2}
\let\VANthebibliography\thebibliography
\def\thebibliography{\DeclareRobustCommand{\VAN}[3]{##3}\VANthebibliography}

\usepackage{graphicx}	
\usepackage{amsmath}	
\usepackage{anyfontsize}

\newcommand{\gaia}{\textit{Gaia}}
\newcommand{\Teff}{\mbox{$T_{\mathrm{eff}}$}}

\newcommand{\logg}{\mbox{$\log g$}}

\title[DZ WDs with exponentially decaying discs]{Characterizing planetary material accreted by cool helium atmosphere white dwarfs using an exponentially decaying disc model}

\author[M. W. O'Brien et al.]{Mairi~W.~O'Brien,$^{1}$\thanks{E-mail: mairi.obrien1@gmail.com}
Pier-Emmanuel~Tremblay,$^{1}$
Beth~L.~Klein,$^{2}$
Carl~Melis,$^{3}$
Detlev~Koester,$^{4}$
\newauthor 
Andrew~M.~Buchan,$^{1}$
Dimitri~Veras,$^{1,5,6}$
and Alexandra~E.~Doyle$^{7}$
\\
$^{1}$ Department of Physics, University of Warwick, Coventry CV4 7AL, UK \\
$^{2}$ Department of Physics and Astronomy, University of California, Los Angeles, CA 90095-1562, USA \\
$^{3}$ Department of Astronomy \& Astrophysics, University of California, San Diego, CA 92093-0424, USA \\
$^{4}$ Institut f\"ur Theoretische Physik und Astrophysik, University of Kiel, 24098 Kiel, Germany \\
$^{5}$ Centre for Exoplanets and Habitability, University of Warwick, Coventry CV4 7AL, UK \\
$^{6}$ Centre for Space Domain Awareness, University of Warwick, Coventry CV4 7AL, UK \\
$^{7}$ Department of Earth, Planetary, and Space Sciences, University of California, Los Angeles, CA 90095, USA \\
}

\date{Accepted 2025 March 07. Received 2025 February 28; in original form 2025 February 10}

\pubyear{2025}

\begin{document}
\label{firstpage}
\pagerange{\pageref{firstpage}--\pageref{lastpage}}
\maketitle

\begin{abstract}
We present Keck High Resolution Echelle Spectrometer (HIRES) observations and model atmosphere analysis for two nearby, cool, helium-dominated atmosphere white dwarfs that have been polluted by accretion: WD\,J1927$-$0355 and WD\,J2141$-$3300. Detected elements common to both white dwarfs are Mg, Ca, Ti, Cr, Fe, and Ni, with additional detections of Na, Al, Si and Sr in WD\,J2141$-$3300. We present an approach for inferring the composition of the accreted material, by adopting a physically motivated model in which the mass accretion rate decays exponentially with time, which provides constraints on the time since the start of the accretion event. The accretion events were most likely to have began at least 1\,Myr ago, however the characteristic disc lifetime could not be constrained due to degeneracies. Both white dwarfs were found to have accreted bulk planetary material with compositions similar to that of both bulk Earth and chondritic meteorites. The parent bodies causing pollution in both cases were inferred to be the mass of a small moon or dwarf planet.
\end{abstract}



\section{Introduction}

In recent years, scientific and instrumental improvements have enabled the precise characterisation and interpretation of the atmospheric composition of exoplanets \citep[see reviews by][]{Madhusudhan2019,Kempton2024}. In contrast, much less is known about the composition of exoplanet interiors \citep[see chapter by][]{Noack2024}. Metals visible in the atmospheres of white dwarfs cooler than 20\,000\,K imply accretion of planetary material, caused by the tidal disruption of planetary bodies that are accreted by the white dwarf host \citep{Jura2003,Farihi2009,Malamud2020,Brouwers2022}, as radiative levitation effects are negligible at these low temperatures. Photospheric metal lines have been detected in the atmospheres of over 1700 white dwarfs \citep{Coutu2019,Williams2024}, and provide a unique insight into the interior composition of exoplanets. Most known planetary systems around main-sequence stars will eventually end up orbiting a white dwarf, and thus the study of current white dwarf planetary systems should on average reproduce main sequence exoplanet properties.

Tidally disrupted planetary bodies produce circumstellar dust discs before being accreted by their white dwarf host \citep{Debes2002,Jura2003,Veras2014}. Dust discs have been observed in 1\,--\,3\,per\,cent of metal polluted white dwarfs \citep{Farihi2009} via the detection of an excess at infrared wavelengths (e.g. \citealt{Zuckerman1987,Jura2007b,Jura2007a,vonHippel2007,Farihi2016,Wilson2019,Xu2020,Lai2021}). These discs should be commonplace around polluted white dwarfs, however most white dwarf dust discs emit insufficient flux to be detected by infrared facilities such as \textit{Spitzer} \citep{Rocchetto2015}. An even smaller number of white dwarfs are observed to harbour a gas component in their disc, which reveals itself through gaseous emission features in optical white dwarf spectra (e.g. \citealt{Gaensicke2006,Gaensicke2007,Gaensicke2008,Dennihy2020,Melis2020,Gentile2021_gas,Rogers2024}). Another small subset of white dwarfs display periodic transits caused by eclipsing planetary material (e.g. \citealt{Vanderburg2015,Xu2016,Vanderbosch2020,Guidry2021,Robert2024}).

Most polluted white dwarfs in the solar neighbourhood have accreted rocky material (e.g. \citealt{Zuckerman2007,Klein2010,Koester2014,Harrison2021,Trierweiler2023,Doyle2023,Swan2023}), the compositions of which typically correspond to chondiritic meteorite or bulk Earth compositions. White dwarfs have also been observed to have accreted core, crust or mantle-rich fragments of a differentiated planetesimal (e.g. \citealt{Zuckerman2011,Melis2011,Gaensicke2012,Melis2017,Hollands2018_dz,Hollands2021,Buchan2022}). There are additionally some detections of an excess of volatile species and hydrogen, possibly due to the accretion of water-rich or icy planetary bodies (e.g. \citealt{Farihi2013,Raddi2015,Xu2017,Xu2019,Hoskin2020,Klein2021}). Many detections of volatile species, especially oxygen, have relied on ultraviolet spectroscopy. However, without ultraviolet data, a limit can still be placed on the accreted abundance of oxygen for optical spectroscopy by assuming a stoichiometric balance with all of the detected rock-forming elements in the white dwarf atmosphere, which has been evidenced in studies where oxygen lines are available \citep{Klein2010,Doyle2023,Rogers2024b}.

Accreted metals sink through the atmosphere of the white dwarf on timescales which vary as a function of atmospheric composition and temperature. In the atmosphere of a metal polluted helium-rich atmosphere white dwarf (typically spectral type DZ or DBZ), metals can have settling timescales of the order of Myrs, such that metals remain visible in the atmosphere long after an accretion event has ceased (\citealt{Paquette1986,Dufour2007_dz,Koester2011,Hollands2017,Hollands2018_dz,Swan2023}). Each metal sinks out of the underlying convection zone at a different rate, and therefore the relative fractions of elements observed in the atmosphere may not reflect the composition of the accreted parent body, depending on the phase of accretion at the time of observation \citep{Koester2009}. The settling timescales of metals in cool DZ and DBZ white dwarfs are comparable to the lifetime of dust discs, which are expected to last on the 
order of Myrs before becoming depleted \citep{Girven2012,Veras2020,Cunningham2021}. 

White dwarf discs are often assumed to accrete at a constant accretion rate until the disc is depleted, at which point accretion ceases \citep{Koester2009}. The theory behind the balance between accretion and diffusion is described in detail in \citet{Dupuis1993}. \citet{Jura2009} instead considered an exponentially decaying accretion rate where the disc lifetime is the characteristic $e$-folding  timescale. \citet{Lodato2008} motivates such a choice as they considered the physics of a disc for which viscosity was proportional to radius, accreting onto a star, and found that the mass accretion rate decayed exponentially as a function of time, and that the disc evolved over a viscous timescale. Following a tidal disruption event, planetary debris is expected to spread and evolve dynamically and the accretion rate should decrease over time as material is depleted from the disc. In contrast, the constant accretion rate model assumes that a steady supply of material is being accreted over an arbitrarily long time.

In this work, we present analysis of optical Keck High Resolution Echelle Spectrometer (HIRES; \citealt{Vogt1994}) spectra of two DZ white dwarfs that were published as part of the volume-limited 40\,pc white dwarf sample \citep{OBrien2024}: WD\,J192743.10$-$035555.23 and WD\,J214157.57$-$330029.80 (WD\,2138$-$332). Throughout the paper, we refer to these white dwarfs by their shorthand WD\,Jhhmm\,$\pm$\,ddmm names: WD\,J1927$-$0355 and WD\,J2141$-$3300. At distances of 24\,pc and 16\,pc respectively, these white dwarfs are some of the most heavily polluted within the local solar neighbourhood, with six metals detected in the atmosphere of WD\,J1927$-$0355 and ten in the atmosphere of WD\,J2141$-$3300. We applied an exponentially decaying disc model from \citet{Jura2009} to the metal abundances of these white dwarfs in order to constrain the composition of the parent bodies that were accreted by both white dwarfs. The \citet{Jura2009} disc model has been adopted in only a handful of white dwarf studies \citep{Jura2010,Doyle2020,Doyle2021,Trierweiler2022}, and this work is the first to treat the disc lifetime as a free parameter when determining the composition of accreted material in observed white dwarfs with this model. 


\section{Observations and Stellar Modelling}

\begin{table*}
    \centering
    \begin{tabular}{llllllll}
        \hline
         WD\,J Name  & WD\,J Name &\textit{Gaia} DR3 ID&  Observation &  Collimator&  Wavelength&  Nominal& Exposure\\
           & (shortened) &&  Date (UT)&  &  Range (\AA)&  Resolution& (s)\\
         \hline
         192743.10$-$035555.23   & 1927$-$0355 &4213409341688406912&  2019 July 07&  Blue&  3050$-$5940&  35\,800& 3000\\
           & &&  2019 July 07&  Blue&  3130$-$5940&  35\,800& 3000\\
           & && 2019 September 07& Red& 4690$-$9140& 35\,800&2100\\
         214157.57$-$330029.80  & 2141$-$3300&6592315723192176896& 2008 August 06& Blue& 3050$-$5940& 23\,800&1500 $\times$ 2\\
           & && 2008 August 07& Blue& 3130$-$5940& 35\,800&1500 $\times$ 2\\
           & && 2008 November 14& Red& 4690$-$9150& 35\,800&1800 $\times$ 2\\
          \hline
    \end{tabular}
    \caption{Keck HIRES observation details.}
    \label{tab:observation_details}
\end{table*}

\begin{figure*}
    \centering
	\includegraphics[width=\textwidth]{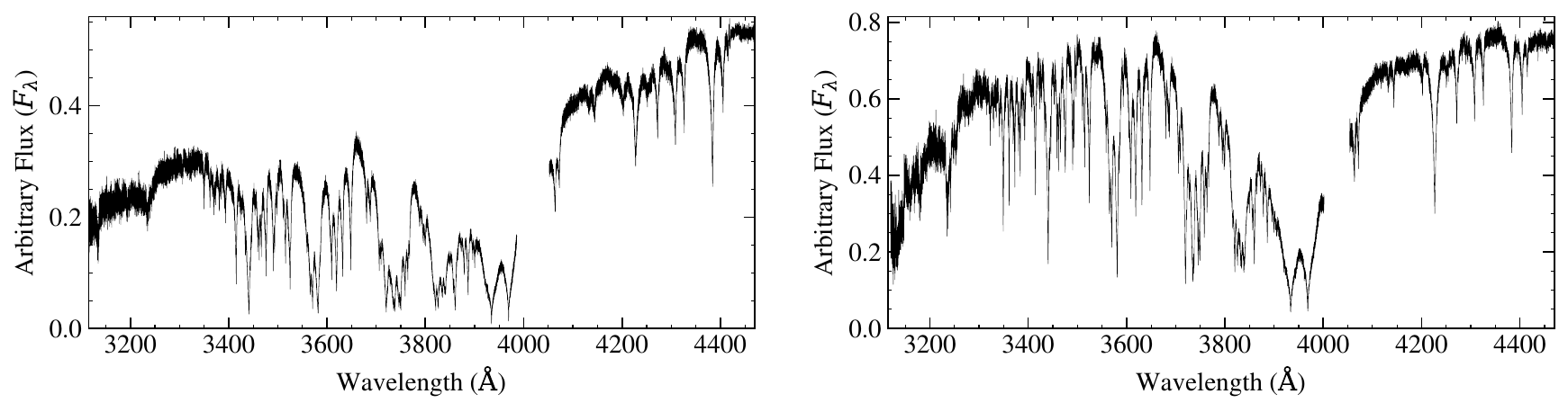}
	\caption{Combined, echelle-order-merged, blue portions of the HIRES spectra of WD\,J1927$-$0355 (left) and WD\,J2141$-$3300 (right). The continua and slopes of the spectra are not well flux calibrated, and therefore the spectra do not accurately depict the true spectral shape for each star.}
    \label{fig:both_WDs_blue_no_model}
\end{figure*}

Hundreds of thousands of new white dwarf candidates were identified in \gaia\ Data Release 2, including WD\,J1927$-$0355 \citep{Gentile2019}. The Kast spectrograph on the Shane 3\,m telescope at the Lick Observatory confirmed its white dwarf status and enabled a DZ classification, and the star was then followed up at higher resolution with HIRES in order to identify extra metal species and hydrogen. WD\,J2141$-$3300 was identified as a DZ white dwarf by \citet{Subasavage2007}. \citet{Bagnulo2019} confirmed that WD\,J2141$-$3300 harbours a weak magnetic field via spectropolarimetric measurements, and \citet{Bagnulo2024a} observed that the equivalent widths of each of its spectral features vary over its rotation period.

Details of the observations analysed in this work are presented in Table~\ref{tab:observation_details}. Reduced and merged HIRES spectra for both white dwarfs are presented in Fig.~\ref{fig:both_WDs_blue_no_model}. The HIRES data of WD\,J1927$-$0355 were reduced using the \href{https://sites.astro.caltech.edu/~tb/makee/}{MAKEE} pipeline. The HIRES data of WD\,J2141$-$3300 were reduced using \texttt{PyRAF} \citep{STScI2012}, following \citet{Klein2010}. Continuum fits of the calibration stars BD+28\,4211, G191$-$B2B, and EGGR\,131, were used to model and divide out the instrumental response function. Procedures described in \citet{Klein2011} were used to align, trim, and merge the echelle orders in order to generate the spectra.

The HIRES spectra extend to the very blue end of the optical ($\approx$\,3100\,\AA) at high signal to noise, enabling the detection of many metal lines. In the spectra of WD\,J1927$-$0355, we identified lines originating from six individual elements: Mg, Ca, Ti, Cr, Fe, and Ni. In the spectra of WD\,J2141$-$3300, we identified Na, Mg, Al, Si, Ca, Ti, Cr, Fe, Ni, and Sr lines. More than one line was detected for each element, aside from Si\,\textsc{i} in WD\,J2141$-$3300, for which there was only one line at 3905.65\,\AA. We identified H$\alpha$ and weak H$\beta$ features in the spectra of both white dwarfs. 

Synthetic spectra for the two white dwarfs were computed using white dwarf atmosphere models. The basic input data and methods are described in \citet{Koester2010}, but the code has been considerably improved regarding the equation of state, 
absorption coefficients, and line broadening theories (see e.g. \citealt{Hollands2017,Gaensicke2018,Wilson2019,Elms2022}). In the subsequent envelope modelling, there were two options: one where convective overshoot was switched on \citep{Kupka2018,Cunningham2019}, meaning its pressure scale height was set equal to one, and a second option with no convective overshoot. We chose to use the output where convective overshoot was switched on.


A hybrid fit of spectroscopy and photometry was used to determine the best-fit parameters of these stars. Metal abundances were changed until all strong features were reasonably well reproduced by models. With these abundances a small model grid was calculated, alongside a table with theoretical absolute magnitudes for all available observed photometry. Fitting to the photometry resulted in new values for the best-fit parameters, and this iteration process was repeated three times to provide a final converged \Teff\ and \logg\ as well as metal abundances.

The best-fit parameters for both white dwarfs are presented in Table~\ref{tab:metal_abundances}. The uncertainties on the number abundances quoted in Table~\ref{tab:metal_abundances} are estimates of the uncertainties derived from spectroscopic fitting, and the uncertainties of \Teff\ and \logg\ were not incorporated in their determination. The influence of these additional sources of uncertainty on the relative abundances is minor. In \citet{OBrien2024}, a typographical error resulted in an incorrect value being reported for $\log$(Mg/He) for WD\,J1927$-$0355. The correct abundance is $-7.80$, as shown in Table~\ref{tab:metal_abundances}, and this correction does not impact any conclusions or analysis in \citet{OBrien2024}. We have also updated the upper limits of copper and cobalt for both white dwarfs in Table~\ref{tab:metal_abundances}, providing less stringent values in order to prevent over-interpretation of these limits. The settling timescales and masses of all elements in the convection zone at the time of observation are presented in Table~\ref{tab:sinking_timescales}.

\begin{figure}
    \centering
	\includegraphics[width=\columnwidth]{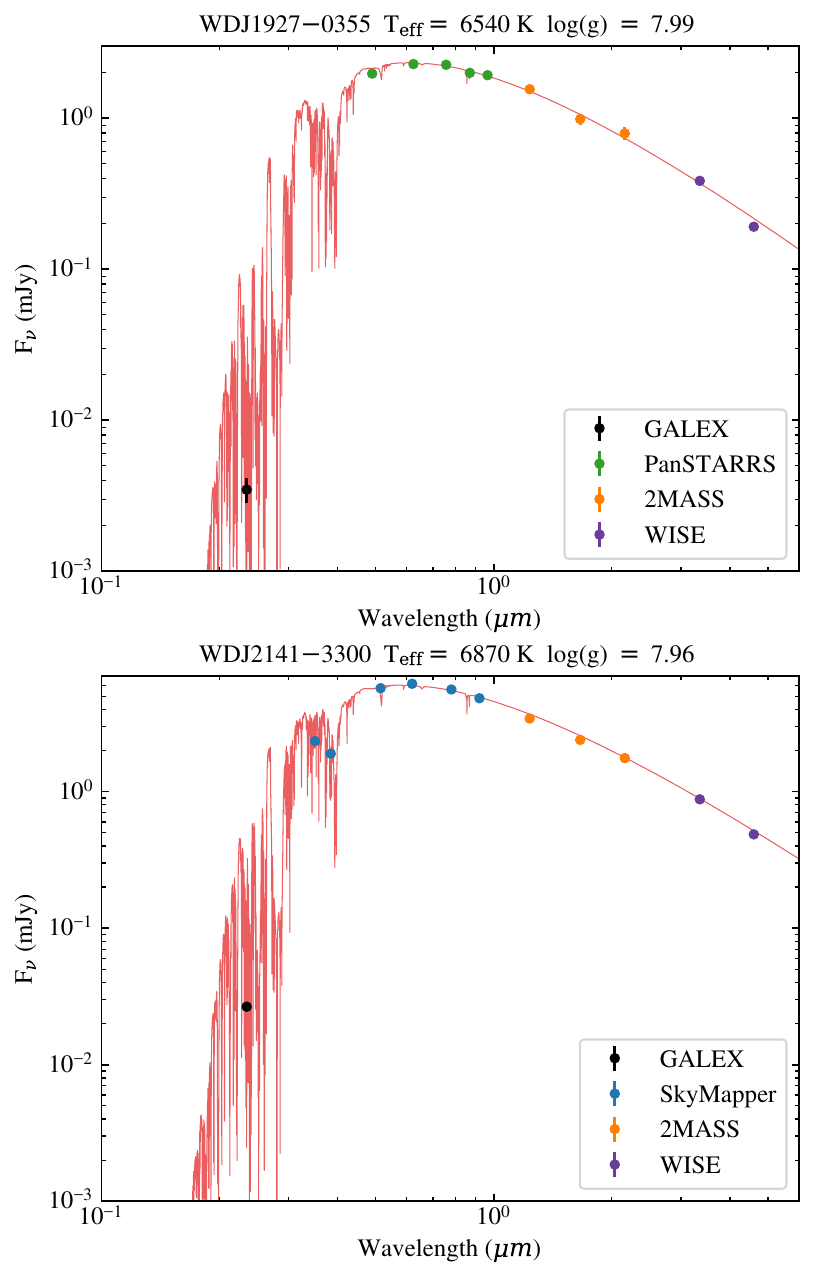}
	\caption{Best-fit model to the metal lines in the HIRES spectra and catalogue photometry for both WD\,J1927$-$0355 (top) and WD\,J2141$-$3300 (bottom). The model was calculated using code from \citet{Koester2010}.}
    \label{fig:both_SED}
\end{figure}

In Fig.~\ref{fig:both_SED}, the best-fit models are plotted alongside GALEX $NUV$ \citep{GALEX2005}, 2MASS $JHK_\mathrm{s}$ \citep{2MASS2006} and WISE $W1$, $W2$ photometry \citep{WISE2010}, plus Pan-STARRS $grizy$ \citep{PanSTARRS2016} for WD\,J1927$-$0355 and SkyMapper $uvgriz$ \citep{SkyMAPPER2005} for WD\,J2141$-$3300. In Table~\ref{tab:metal_abundances}, we assumed that the distance is the inverse of the \gaia\ Data Release 3 parallax, which is appropriate for these stars since their parallax error is less than 0.2\,per\,cent. We used these distances to scale the model spectra to the photometry. Fig.~\ref{fig:both_SED} shows that neither white dwarf displays an excess indicative of a dust disc above the white dwarf flux in their 2MASS and WISE infrared photometry. \citet{Girven2012} presented \textit{Spitzer} IRS peak-up imaging of WD\,J2141$-$3300, detecting a 3\,$\sigma$ flux excess at 15.6\,$\mu$m, but mentioned that this excess could be caused by a background galaxy. They noted that a dust disc emitting at such long wavelengths must be cool and probably outside of the region where material would be accreted by the star, meaning it might not be the cause of the current observed pollution.

\begin{figure*}
    \centering
	\includegraphics[width=\textwidth]{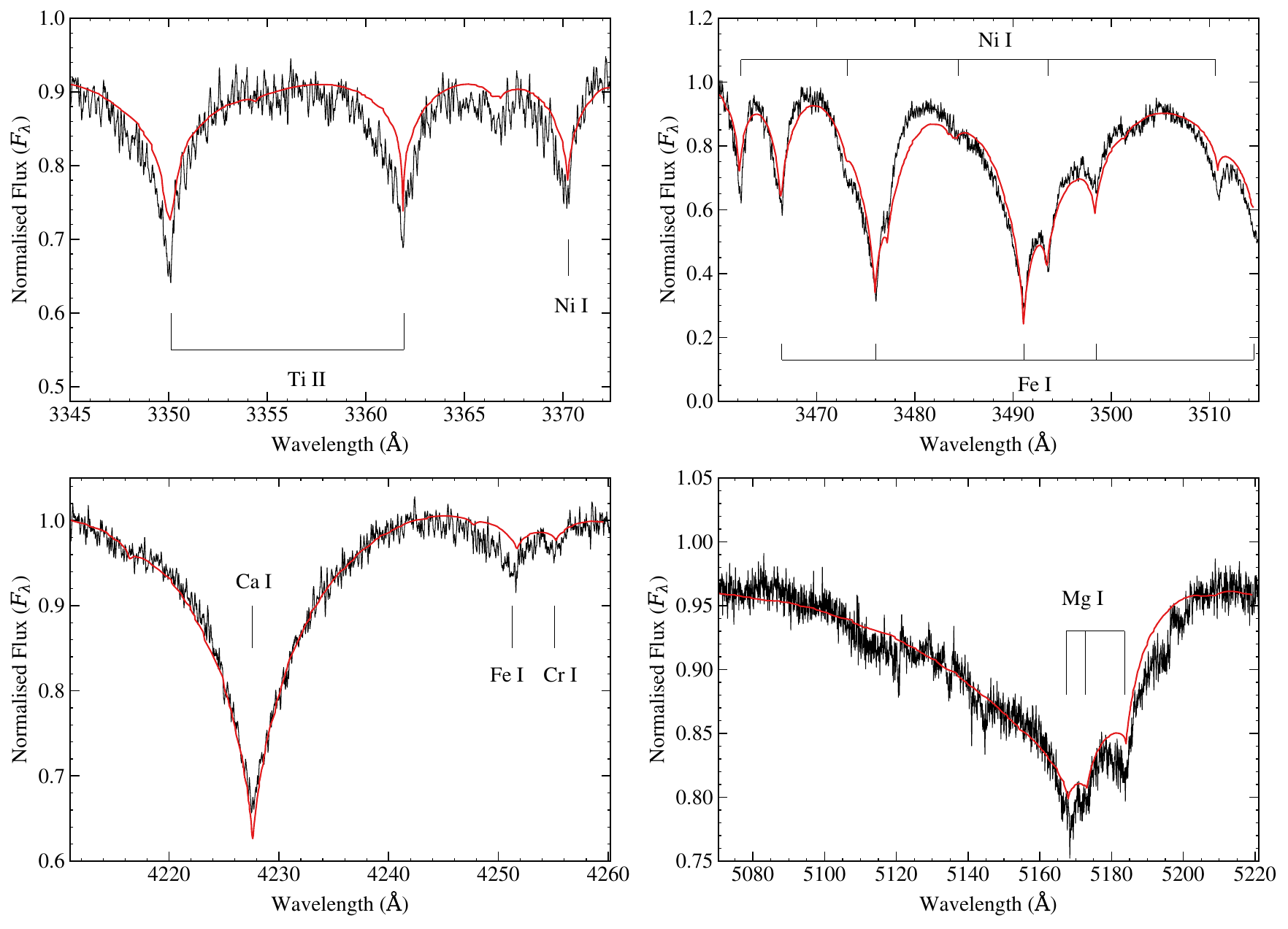}
	\caption{A selection of Ti\,\textsc{ii}, Cr\,\textsc{i}, Ca\,\textsc{i}, Mg\,\textsc{i}, Ni\,\textsc{i} and Fe\,\textsc{i} lines in the HIRES spectrum of WD\,J1927$-$0355, with the best-fit model overplotted in red. The data were smoothed with a five-point boxcar average and the model spectra were convolved with a Gaussian function. The wavelengths of the models were shifted to match the best-fit radial velocity shift of the data. The spectra and models were normalised to the continuum level and detrended for improved visual clarity. There are many lines of each metal throughout the spectrum.}
    \label{fig:wdj1927_lines}
\end{figure*}

\begin{figure*}
    \centering
	\includegraphics[width=\textwidth]{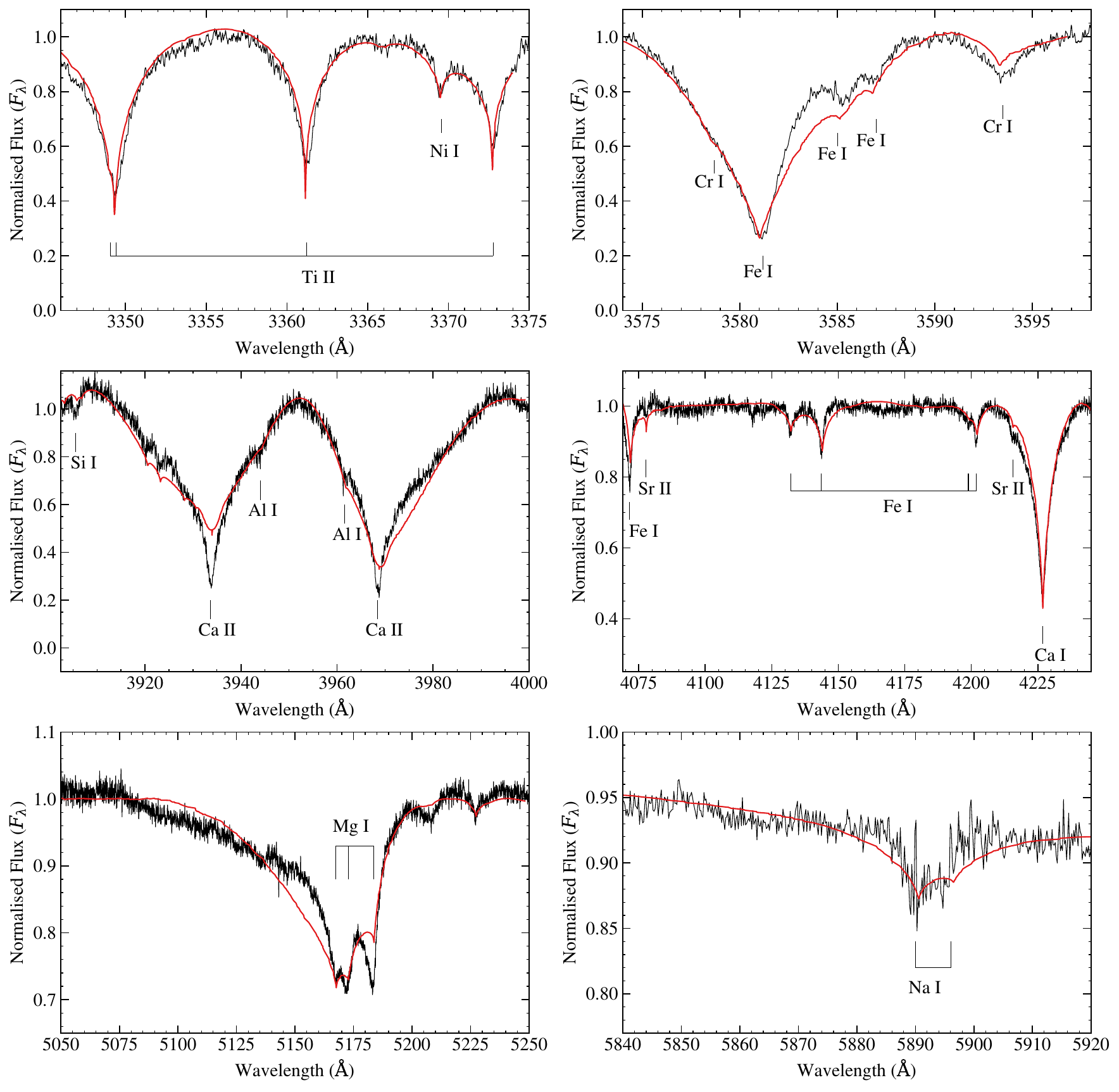}
	\caption{A selection of Ti\,\textsc{ii}, Cr\,\textsc{i}, Ca\,\textsc{i}, Ca\,\textsc{ii}, Mg\,\textsc{i}, Na\,\textsc{i}, Si\,\textsc{i}, Al\,\textsc{i}, Sr\,\textsc{ii}, and Fe\,\textsc{i} lines in the HIRES spectrum of WD\,J2141$-$3300, with the best-fit model overplotted in red. The data were smoothed with a five-point boxcar average and the model spectra were convolved with a Gaussian function. The wavelengths of the models were shifted to match the best-fit radial velocity shift of the data. The spectra and models were normalised to the continuum level and detrended for improved visual clarity. There are many lines of each metal throughout the spectrum. Some portions of the spectrum are poorly calibrated despite normalisation and detrending. Some fits are also poor due to Zeeman splitting of metal lines caused by the magnetic field at WD\,J2141$-$3300, which was not incorporated into the models.}
    \label{fig:wdj2141_lines}
\end{figure*}

Fits to individual metal lines are shown in Fig.~\ref{fig:wdj1927_lines} for WD\,J1927$-$0355 and Fig.~\ref{fig:wdj2141_lines} for WD\,J2141$-$3300. The data in these plots were smoothed with a five-point boxcar average and the model spectra were convolved with a Gaussian function according to the resolution of the observations listed in Table~\ref{tab:observation_details}. The wavelengths of the models were shifted to match the best-fit radial velocity shift of the data in the heliocentric frame of rest. The spectra and models were normalised to the continuum level and detrended for improved visual clarity. The reduction of echelle spectra is notoriously challenging, and there were substantial issues with the shapes and slopes of the spectra following reduction. The blue parts of the spectra for both white dwarfs have few points which can be considered as ‘continuum’, and therefore normalising the spectrum to match the model was not straightforward. Less than optimal visual fits to some lines in the spectrum of WD\,J2141$-$3300 are due to weak Zeeman splitting that was not accounted for in the models. The continuum mismatch near Sr\,\textsc{ii} in Fig.~\ref{fig:wdj2141_lines} is an artifact of the normalisation procedure, and the strontium abundance was determined with a continuum range that was local to the line. At the time of publication, strontium has been detected in the atmosphere of just three other white dwarfs: GD\,362 \citep{Zuckerman2007}, WD\,0446$-$255 \citep{Swan2019}, and 2MASS J0916$-$4215 \citep{Vennes2024}. Plots showing the full fits to the HIRES spectra can be found in Fig. A2 of the supplementary data of \citet{OBrien2024}.

\begin{figure}
    \centering
	\includegraphics[width=\columnwidth]{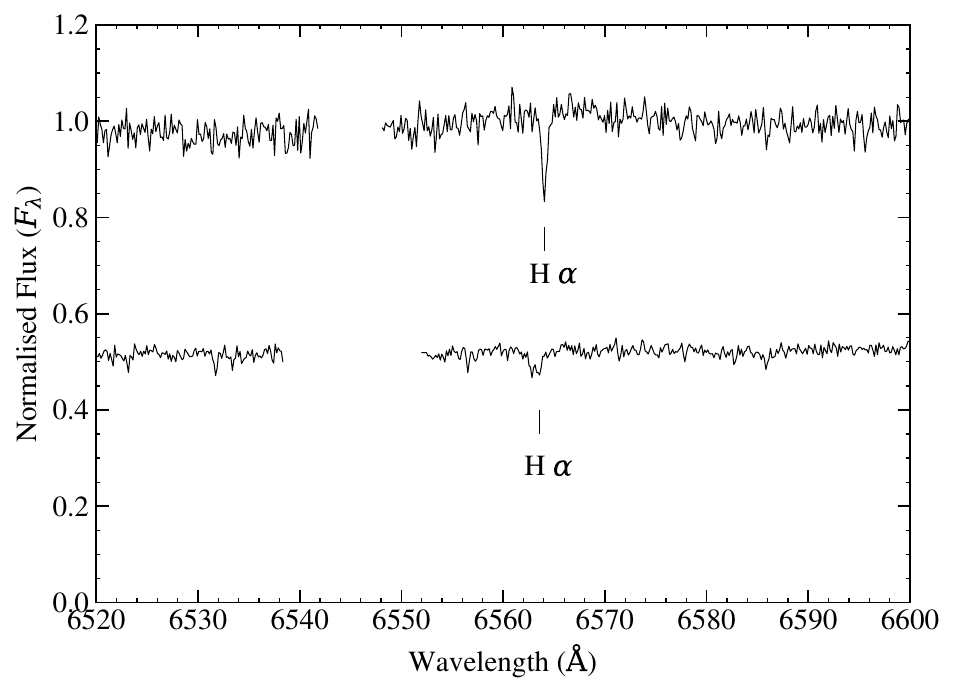}
	\caption{Normalised and binned HIRES spectra of WD\,J1927$-$0355 (top) and WD\,J2141$-$3300 (bottom) in the region around the H$\alpha$ line. A constant offset has been applied to the flux of WD\,J2141$-$3300, for visual clarity.}
    \label{fig:halpha}
\end{figure}

Figure~\ref{fig:halpha} shows H$\alpha$ detections with non-local thermodynamic equilibrium line cores for both white dwarfs. The equivalent widths of the H$\alpha$ lines are 170\,m\AA\ for WD\,J1927$-$0355 and 30\,m\AA\ for WD\,J2141$-$3300, the latter of which also shows Zeeman splitting of the line core. The temperature of a hydrogen-rich atmosphere white dwarf model required to match the H$\alpha$ equivalent widths in Fig.~\ref{fig:halpha} would be $<$\,5000\,K, which is discrepant with the photometric effective temperatures for both white dwarfs, which are 6500\,$-$\,7000\,K. Therefore, both white dwarfs must have helium-rich atmospheres with some hydrogen. It was not possible to match both the H$\alpha$ features and the shapes of the metal lines using current helium-rich atmosphere white dwarf models. Therefore, a standard ratio of $\log$(H/He)\,$=$\,$-3.5$ was adopted for the models of both white dwarfs, in accordance with measured hydrogen abundances in similar DZA white dwarfs \citep{Hollands2017, Coutu2019}. Models with this composition fit the metal line shapes well, but produce a very broad and shallow H$\alpha$ feature compared to the narrow, sharp H$\alpha$ lines observed in the spectra. Unexpectedly sharp and narrow H$\alpha$ features have also been observed in other cool DZA white dwarfs \citep{OBrien2023}.

\begin{table}
	\centering
        \label{tab:metal_abundances}
        \begin{tabular}{lll}
                \hline
                Parameter & WD\,J1927$-$0355 & WD\,J2141$-$3300 \\
                \hline
                Spectral Type& DZA&DZAH\\
                \Teff\ [K] & 6540 (150) & 6870 (150) \\
                \logg\ [cm s$^{-2}$] & 7.99 (0.04) & 7.96 (0.04) \\
                Distance [pc]& 23.85 (0.04)&16.111 (0.006)\\
                $*$ log(H/He) & $-$3.5 & $-$3.5 \\
                log(Na/He) & $<-$9.4& $-$9.2 (0.2) \\
                log(Mg/He) & $-$7.80 (0.15) & $-$7.50 (0.15) \\
                log(Al/He) & $<-$8.6 & $-$8.5 (0.3) \\
                log(Si/He) & $<-$7.6 & $-$7.2 (0.2) \\
                log(Ca/He) & $-$9.1 (0.1) & $-$8.9 (0.1) \\
                log(Sc/He) & $<-$11.7 & $<-$11.7 \\
                log(Ti/He) & $-$10.7 (0.1) & $-$10.0  (0.1) \\
                log(V/He) & $<-$10.5 & $<-$10.4 \\
                log(Cr/He) & $-$10.2 (0.2) & $-$10.0 (0.2) \\
                log(Mn/He) & $<-$9.2 & $<-$9.3 \\
                log(Fe/He) & $-$8.0 (0.1) & $-$8.2 (0.1) \\
                log(Co/He) & $<-$9.8& $<-$10.0\\
                log(Ni/He) & $-$9.3 (0.2) & $-$9.2 (0.2) \\
                log(Cu/He) & $<-$11.5& $<-$11.2\\
                log(Sr/He) & $<-$12.3 & $-$12.1 (0.3) \\
                log(Ba/He) & $<-$12.2 & $<-$12.2 \\
                \hline
        \end{tabular}\\
        \caption{Abundances and upper limits of elements determined from combined fitting of spectra and photometry with \citet{Koester2010} models. $*$ indicates that the abundance was fixed, not measured. Distances were determined from inverting \gaia\ Data Release 3 parallaxes, and distance errors incorporated the \gaia\ astrometric excess noise.}
\end{table} 

\begin{table*}
    \centering
    \begin{tabular}{llllll}
    \hline
         Atomic&  Element&  $\log(t\textsubscript{set}$/yr)&  $M\textsubscript{CVZ}$ / 10$^{21}$\,g& $\log(t\textsubscript{set}$/yr)&$M\textsubscript{CVZ}$ / 10$^{21}$\,g\\
 Number& Z& WD\,J1927$-$0355& WD\,J1927$-$0355& WD\,J2141$-$3300&WD\,J2141$-$3300\\
         \hline        
         \vspace{1mm}
         1&  $*$\,H& $-$ & 1071  & $-$ & 1259 \\
                \vspace{1mm}
         11&  Na&  6.55&  $<$ 0.03& 6.62&0.06$^{+0.03}_{-0.02}$\\
                 \vspace{1mm}
         12&  Mg&  6.55&  1.3$^{+0.5}_{-0.4}$ & 6.62&3$^{+1}_{-1}$\\
                 \vspace{1mm}
         13&  Al&  6.50&  $<$ 0.23 & 6.57&0.3$^{+0.3}_{-0.2}$\\
                 \vspace{1mm}
         14&  Si&  6.51&  $<$ 2.37 & 6.58&7$^{+4}_{-3}$\\
                 \vspace{1mm}
         20&  Ca&  6.37&  0.11$^{+0.03}_{-0.02}$ & 6.45&0.20$^{+0.05}_{-0.04}$\\
                 \vspace{1mm}
         21&  Sc&  6.31&  $<$ 0.0003 & 6.38&$<$ 0.0004\\
                 \vspace{1mm}
         22&  Ti&  6.28&  0.0032$^{+0.0008}_{-0.0007}$ & 6.35&0.019$^{+0.005}_{-0.004}$\\
                 \vspace{1mm}
         23&  V&  6.25&  $<$ 0.005 & 6.33&$<$ 0.008\\
                 \vspace{1mm}
        24& Cr& 6.25& 0.011$^{+0.006}_{-0.004}$ & 6.33&0.021$^{+0.012}_{-0.008}$\\
                \vspace{1mm}
        25& Mn& 6.22& $<$ 0.12 & 6.30&$<$ 0.11\\
                \vspace{1mm}
        26& Fe& 6.23& 1.9$^{+0.5}_{-0.4}$ & 6.30&1.4$^{+0.4}_{-0.3}$\\
                \vspace{1mm}
        27& Co& 6.20& $<$ 0.003 & 6.28&$<$ 0.02\\
                \vspace{1mm}
        28& Ni& 6.22& 0.10$^{+0.06}_{-0.04}$ & 6.29&0.15$^{+0.09}_{-0.05}$\\
                \vspace{1mm}
        29& Cu& 6.17& $<$ 0.0007 & 6.25&$<$ 0.0015\\
                \vspace{1mm}
        38& Sr& 6.04& $<$ 0.0001 & 6.12&0.0003$^{+0.0003}_{-0.0001}$\\
                \vspace{1mm}
        56& Ba& 5.87& $<$ 0.0003 & 5.95&$<$ 0.0003\\
        \hline
    \end{tabular}
    \caption{Gravitational settling times and masses of heavy elements in the convection zone for WD\,J1927$-$0355 and WD\,J2141$-$3300, for the models with overshoot. The total mass of heavy elements (i.e. excluding hydrogen) detected in the convection zone is 6.15$\times$\,10$^{21}$\,g for WD\,J1927$-$0355 and 1.23$\times$\,10$^{22}$\,g for WD\,J2141$-$3300. The $*$ indicates that the hydrogen abundance was fixed at log(H/He)\,$=$\,$-$3.5.}
    \label{tab:sinking_timescales}
\end{table*}

WD\,J2141$-$3300 has an average mean magnetic field modulus of 50\,$\pm$\,10\,kG \citep{Bagnulo2024b} and rotates with a period of 6.2\,hours \citep{Hernandez2024, Farihi2024}. WD\,J2141$-$3300 was observed to have varying equivalent widths of metal lines within its rotation period \citep{Bagnulo2024b}. A hypothesis for this variation is that there is an inhomogeneous chemical composition at the white dwarf photosphere which changes in visibility as the star rotates \citep{Bagnulo2024b, Bagnulo2024a}. \citet{Bagnulo2024b} found that the measured metal abundances varied over the rotation period of the white dwarf, but that all the abundances varied in phase. Therefore, the ratios of metal abundances should remain roughly constant with time. The analysis presented in this work relies only on abundance ratios and the variation in line strength should not significantly affect our results. 

\section{Methods}
\label{sec:methods}

\subsection{Exponentially decaying disc model}

\begin{figure}
    \centering
	\includegraphics[width=\columnwidth]{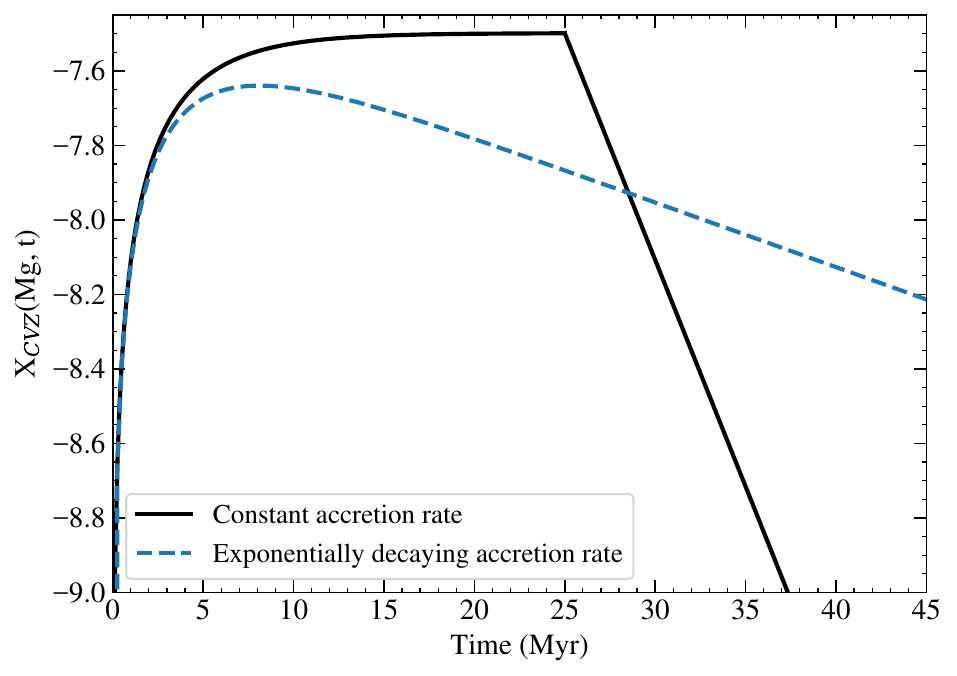}
	\caption{A demonstration of the mass abundance (X\textsubscript{CVZ}) of Mg in the convection zone of WD\,J1927$-$0355 for the constant accretion rate model \citep{Dupuis1993,Koester2009} and for the \citet{Jura2009} exponentially decaying model with a 25\,Myr disc $e$-folding characteristic timescale. The accretion was switched off at 25\,Myr for the constant accretion rate model scenario, for illustrative purposes. $\dot M\textsubscript{acc}(t=0)$ for the exponentially decaying accretion rate model was set equal to $\dot M\textsubscript{acc}$ for the constant accretion rate model.}
    \label{fig:wdj1927_model_comparison}
\end{figure}

For the analysis of the composition of the material accreted by these two white dwarfs, we employed the model of \citet{Jura2009}, which considers a disc for which the accretion rate decays exponentially with a characteristic $e$-folding timescale $t$\textsubscript{disc}. This model is appropriate for a disc in which viscosity is proportional to radius \citep{Lodato2008}, and for a disc which is not replenished and decays over time. Cool helium-atmosphere white dwarfs are the systems for which the choice of disc model has the most substantial effect, since the disc lifetime and settling timescales in these systems are similar orders of magnitude. 

In the \citet{Jura2009} model, the mass accretion rate decays according to
\begin{equation}
    \dot M\textsubscript{acc}(t) = \frac{M\textsubscript{PB}}{t\textsubscript{disc}} e^{-t/t\textsubscript{disc}}~, 
    \label{eq:decaying_acc_rate}
\end{equation}

{\noindent}where $\dot M\textsubscript{acc}(t)$ is the mass accretion rate as a function of time $t$, and $M\textsubscript{PB}$ is the total mass of the parent body. In this context, parent body refers to the total planetary body that gets accreted within the duration of the overall accretion event. 

The mass of a particular element $Z$ observed at time $t$ in the convection zone of the white dwarf, is given by
\begin{equation}
    M\textsubscript{CVZ}(Z, t) = \frac{M\textsubscript{PB}(Z) t\textsubscript{set}(Z)}{t\textsubscript{disc} - t\textsubscript{set}(Z)} \left( e^{-t/t\textsubscript{disc}} - e^{-t/t\textsubscript{set}(Z)} \right)~,
    \label{eq:jura_main}
\end{equation}

{\noindent}in the \citet{Jura2009} model, where $M\textsubscript{CVZ}(Z, t)$ is the mass of element $Z$ in the convection zone of the white dwarf at time $t$, $t$\textsubscript{set}(Z) is the settling timescale for element Z, and $M\textsubscript{PB}(Z)$ is the mass of element $Z$ in the parent body. 

The characteristic $e$-folding timescale of the disc affects the composition and mass of material accreted, but is poorly constrained. \citet{Girven2012} inferred from infrared observations that disc lifetimes could range from 0.03\,Myr to 5\,Myr. With updated models and a larger sample of white dwarfs with infrared excesses from \citet{Farihi2016}, \citet{Cunningham2021} derived a characteristic lifetime range for white dwarf discs of log($t$\textsubscript{disc} / yr)\,$=$\,6.1\,$\pm$\,1.4, which corresponds to a range of 0.05\,Myr to 32\,Myr, and which we adopt as bounds in this work. \citet{Wyatt2014} inferred that disc lifetimes could be as low as 20\,yr, however they mention that pollution could be sustained for longer if there were multiple consecutive accretion events of this duration. For simplicity, however, in this work, we have assumed that the accretion of a single object, that we refer to as the parent body, has produced the pollution. 

Figure~\ref{fig:wdj1927_model_comparison} demonstrates the difference between the exponentially decaying accretion rate model \citep{Jura2009} and the constant accretion rate model \citep{Dupuis1993,Koester2009}. A steady state is never reached in the exponentially decaying model under the conditions demonstrated in Fig.~\ref{fig:wdj1927_model_comparison}, due to the settling timescales of WD\,J1927$-$0355 being of the order of a typical disc lifetime (see Table~\ref{tab:sinking_timescales}). For the cool DZ(A) white dwarfs discussed in this work, a steady state solution cannot be reached for a typical disc lifetime, as $t$\textsubscript{disc}\,$\simeq$\,$t$\textsubscript{set} or $t$\textsubscript{disc}\,$<<$\,$t$\textsubscript{set}, depending on the disc lifetime. However, as noted by \citet{Koester2009}, if $t$\textsubscript{disc}\,$>>$\,$t$\textsubscript{set}, as is the case for hot hydrogen-atmosphere and hot helium-atmosphere white dwarfs, a steady state-like solution can be recovered in the exponentially decaying disc model. 

We rearranged equation~\ref{eq:jura_main} to incorporate the observed mass of each element in the convection zone, as in equation~3 of \citet{Trierweiler2022}
\begin{equation}
    M\textsubscript{PB}(Z, t\textsubscript{elapse}) = \frac{M\textsubscript{CVZ}(Z) (t\textsubscript{disc} - t\textsubscript{set}(Z))}{t\textsubscript{set}(Z) ( e^{-t\textsubscript{elapse}/t\textsubscript{disc}} - e^{-t\textsubscript{elapse}/t\textsubscript{set}(Z)})}~,
    \label{eq:m_pb}
\end{equation}

{\noindent} where $t$\textsubscript{elapse} is the time elapsed since the start of the accretion event. This rearranged equation and re-labelling of  $t$\textsubscript{elapse} enables it to be set as a free parameter, and represents the phase of accretion that we are observing the system to be in. In Section~\ref{sec:results}, we use equation~\ref{eq:m_pb} to infer the parent body composition, given the observed number abundances and therefore convection zone masses at the time of observation, with $t$\textsubscript{elapse} and $t$\textsubscript{disc} set as free parameters. The two-dimensional parameter space of disc lifetime and time of observation provides constraints on both the phase of accretion and the total mass of the accreted material.

\subsection{Comparisons to solar system compositions}

To learn more about the accreted material and the phase of accretion, we compared the composition of the material accreted at different phases of accretion with many solar system abundances. The best-matching compositions imply that the white dwarf was likely to have accreted material of a particular composition, observed at a particular time after the accretion event began.

Number abundances $n(Z)$ in a white dwarf atmosphere can be converted from mass abundances via: 
\begin{equation}
    \log \left( \frac{n(Z\textsubscript{1})}{n(Z\textsubscript{2})} \right) = \log \left( \frac{X(Z\textsubscript{1})}{X(Z\textsubscript{2})} \frac{m(Z\textsubscript{2})}{m(Z\textsubscript{1})} \right)~,
    \label{eq:num_abund}
\end{equation}

{\noindent}where $X(Z)$ is the mass fraction of element $Z$ compared to the total mass, and $m(Z)$ is the atomic mass of element $Z$. 

In the context of solar system abundances, the mass fraction is the percentage composition by mass of element $Z$ in that object. To find the best compositional match to the material accreted by each white dwarf, we used equation~\ref{eq:m_pb} from the \citet{Jura2009} model to calculate the masses of each element in the parent body, and by extension the number abundances in the parent body, for all combinations of $t$\textsubscript{elapse} and $t$\textsubscript{disc} within the parameter space. We then calculated reduced $\chi^2$ values for number abundances propagated through equation~\ref{eq:m_pb}, ratioed to a reference element, and compared them with various solar system compositions. The total inferred parent body mass was not important for the calculations, as it cancelled out in the ratios between each element and the reference element.

The reduced $\chi^2$ ($\chi^2_{\nu}$) is defined as
\begin{equation}
    \chi^2_{\nu} =  \frac{1}{\nu} \sum_{i=1}^{N} \left( \frac{\left(Y\textsubscript{i,WD} - Y\textsubscript{i,SS}\right)^2}{\sigma^2\textsubscript{WD}(Z\textsubscript{i}) + \sigma^2\textsubscript{WD}(Z\textsubscript{ref})} + \frac{1\,-\,S(Y\textsubscript{i,WD})}{\sigma\textsubscript{i}} \right)~,
    \label{eq:chi2}
\end{equation}

{\noindent}where $\nu$ is the number of degrees of freedom (the total number of elements detected in the white dwarf atmosphere, $N$, minus the number of free parameters, $t$\textsubscript{disc} and $t$\textsubscript{elapse}), $Y_i\,=\,\log\left(\frac{n(Z\textsubscript{i})}{n(Z\textsubscript{ref})}\right)$ for the white dwarf or comparison solar system composition, `WD' and `SS' represent the white dwarf and the comparison solar system composition respectively, $n(Z\textsubscript{ref})$ is the abundance of the reference element, which we chose to be magnesium, and $\sigma\textsubscript{WD}(Z\textsubscript{i})$ is the error on the white dwarf number abundance. For any upper limit measurements in the white dwarf for which abundances of a given meteorite class were available, we did not directly use the limit in the $\chi^2_{\nu}$ calculation, but applied a survival function $S(Y\textsubscript{WD})$ to the $\chi^2_{\nu}$ value. The survival function was adapted from \citet{Swan2023} and references therein, and is of the form,

\begin{equation}
    S(Y\textsubscript{WD}) = \frac{1}{2} \left[ 1 - \mathrm{erf} \left( \frac{Y\textsubscript{WD} - Y\textsubscript{SS}}{\sqrt{2} \sigma\textsubscript{i}} \right) \right]~,
    \label{eq:upperlim}
\end{equation}

{\noindent}where the function \textit{erf} is the Gauss error function. Since this function is being used for upper limits, the uncertainty $\sigma\textsubscript{i}$ is set at a standard value of 0.3\,dex, as in \citet{Swan2023}, which corresponds to the largest error on an abundance measurement in Table~\ref{tab:metal_abundances}. For each upper limit, the term $\frac{1\,-\,S(Y\textsubscript{WD})}{\sigma\textsubscript{i}}$ is applied to the upper limits to incorporate a penalty to the $\chi^2_{\nu}$ value if the upper limit lies below the meteorite abundance. The choice of reference element was somewhat arbitrary, but checks showed that other reference elements did not substantially change the regions of parameter space that best matched the meteorite compositions. The bounds on the input free parameters were set at 0.05\,$-$\,32\,Myr for the disc lifetime from \citet{Cunningham2021}, and 0\,$-$\,100\,Myr for the time since the accretion event began. 

The method for determining the most likely composition of the parent body as well as $t$\textsubscript{elapse} and $t$\textsubscript{disc} proceeds as follows: 
\begin{enumerate}
    \item Select a meteorite class and determine the number abundances of all elements in ratio with magnesium. 
    \item Create a 500\,$\times$\,500 log-spaced grid of values of $t$\textsubscript{disc} (0.05\,$-$\,32\,Myr) and $t$\textsubscript{elapse} (0\,$-$\,100\,Myr), which are both free parameters in the disc model.
    \item For all combinations of $t$\textsubscript{disc} and $t$\textsubscript{elapse} in the grid, convert each number abundance observed in the white dwarf convection zone, as well as each upper limit, into a mass in the convection zone.
    \item Propagate the mass of each element in the convection zone through equation~\ref{eq:m_pb} to determine the mass of each element in the parent body at each pair of $t$\textsubscript{disc} and $t$\textsubscript{elapse}. Convert these masses back to number abundances, and ratio with magnesium.
    \item Use equation~\ref{eq:chi2} to determine the $\chi^2_{\nu}$ comparing: the number abundances of all detected elements in the parent body for a combination of $t$\textsubscript{disc} and $t$\textsubscript{elapse}, and the number abundances of all elements in the chosen meteorite class. Treat upper limits slightly differently: instead of calculating $\chi^2_{\nu}$, use a survival function to add a penalty to the total $\chi^2_{\nu}$.
    \item Produce a $\chi^2_{\nu}$ contour plot for all combinations of $t$\textsubscript{disc} and $t$\textsubscript{elapse} to determine if there are degeneracies in the two parameters, i.e. if many combinations provide an almost equally good comparison.
    \item Repeat this process for all meteorite classes or other solar system compositions, e.g. bulk Earth.
\end{enumerate}

We determined $\chi^2_{\nu}$ values as described above for abundances of both white dwarfs compared to the median composition of each major meteorite class provided in the \citet{Nittler2004} database: carbonaceous (C) chondrites, L chondrites, LL chondrites, H chondrites, E chondrites, aubrites, brachinites, eucrites, diogenites, howardites, urelites, mesosiderites, and pallasites. We also compared the white dwarf abundances to the bulk Earth composition \citep{Allegre2001} and Earth's crust \citep{Rudnick2003}. 

\section{Results}
\label{sec:results}

\begin{table}
    \centering
    \begin{tabular}{llll}
    \hline
         Composition &&  Minimum $\chi^2_{\nu}$  &Minimum $\chi^2_{\nu}$  \\
          && 
    WD\,J1927$-$0355 &WD\,J2141$-$3300\\
    \hline
 Bulk Earth&& 0.8&1.8\\
  \hline
 Earth Crust&& 65.2&31.0\\
  \hline
 Chondrites&C& 1.9&1.3\\
  &E& 2.8&2.7\\
  &H& 2.0&2.3\\
 & L& 2.8&1.9\\
 & LL& 3.2&1.8\\
  \hline
 Achondrites& Aubrites& 51.9&20.7\\
 & Brachinites& 3.9&2.3\\
 & Diogenites& 32.1&16.8\\
 & Eucrites& 63.2&24.6\\
 & Howardites& 31.5&10.8\\
 & Urelites& 9.9&8.3\\
  \hline
 Stony-iron& Mesosiderites& 9.4&1.6\\
 & Pallasites& 18.3&42.8\\
 \hline
 \end{tabular}
    \caption{Minimum $\chi^2_{\nu}$ for solar system abundances compared to the abundances and upper limits of elements in WD\,J1927$-$0355 and WD\,J2141$-$3300.}
    \label{tab:min_chi2}
\end{table}

All major meteorite classes were compared to the abundances and upper limits of elements in WD\,J1927$-$0355 and WD\,J2141$-$3300, propagated through equation~\ref{eq:m_pb} with $t$\textsubscript{disc} and $t$\textsubscript{elapse} as free parameters. The results for the minimum $\chi^2_{\nu}$ are shown in Table~\ref{tab:min_chi2}. The two free parameters are degenerate, and therefore the minimum $\chi^2_{\nu}$ corresponds to a range of parameter space, as explained below. The current sample size of white dwarfs with abundance measurements of the trace elements scandium, vanadium, cobalt, copper, strontium and barium is too small to assess consistency, and their abundance ratios often show variability. As such, these elements were not used for determining constraints on the phase of accretion, even when they were available in datasets including the \citet{Allegre2001} bulk Earth composition. 

\begin{figure}
    \centering
	\includegraphics[width=\columnwidth]{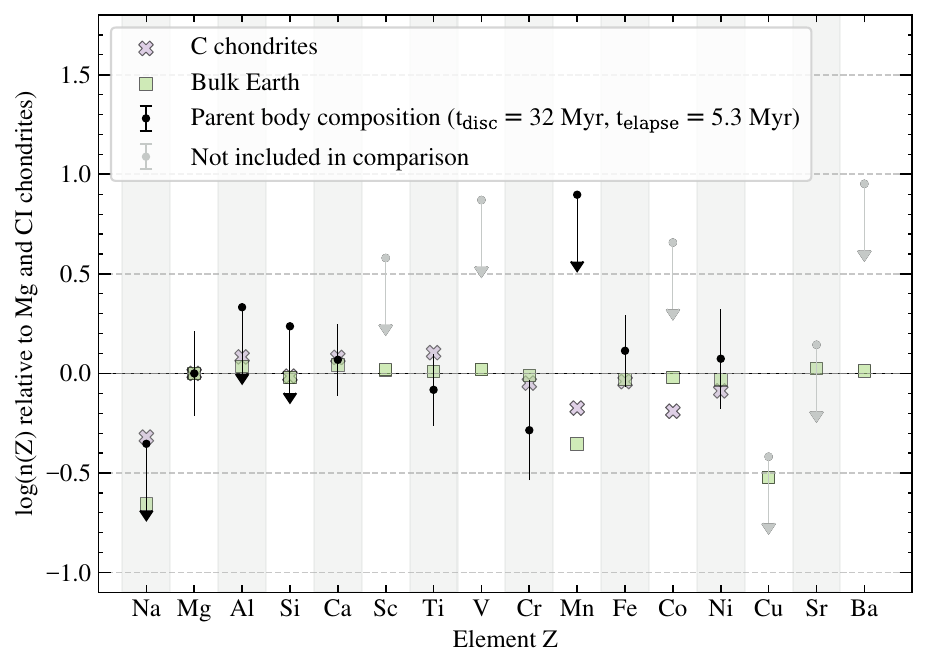}
	\caption{Element abundances for WD\,J1927$-$0355, propagated through the \citet{Jura2009} model for the values of $t$\textsubscript{disc} and $t$\textsubscript{elapse} corresponding to the minimum $\chi^2_{\nu}$, normalised to Mg and CI chondrites, are shown in black. Elements are ordered from left to right in order of increasing atomic weight. Bulk Earth abundances normalised to Mg and CI chondrites are shown by square symbols \citep{Allegre2001}. C chondrite abundances normalised to Mg and CI chondrites are shown by cross symbols \citep{Nittler2004}. Upper limits are denoted with a downwards arrow. CI chondrite abundances are from \citet{Lodders2019}. Error bars are propagated using the errors on abundances from Table~\ref{tab:metal_abundances}. Faded (grey) points represent elements that were not considered in the comparison.}
    \label{fig:WDJ1927_comparison_to_compositions}
\end{figure}

\begin{figure}
    \centering
	\includegraphics[width=\columnwidth]{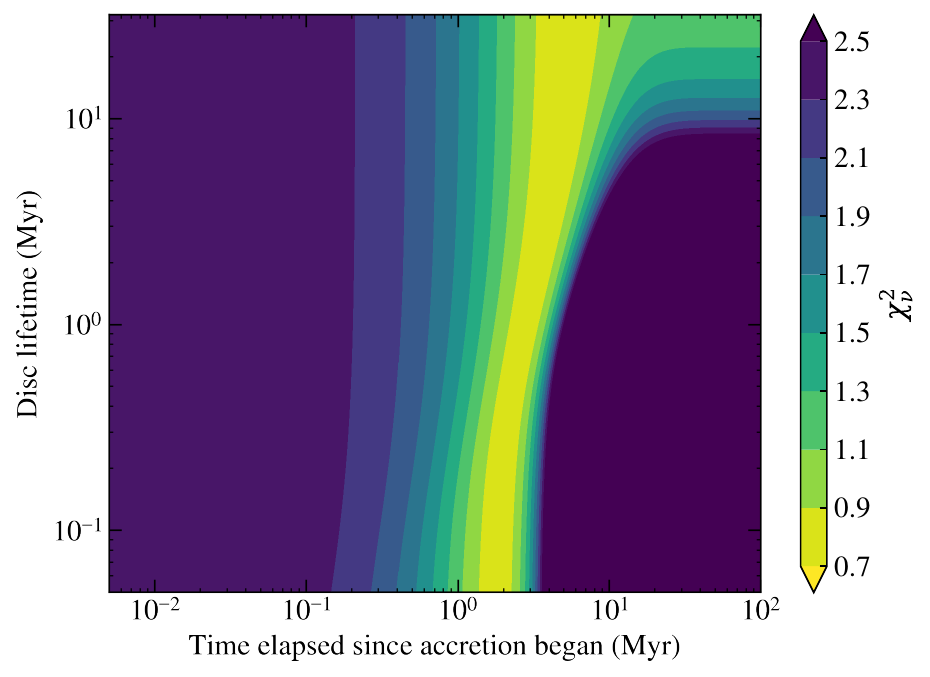}
	\caption{$\chi^2_{\nu}$ contour plot for the abundances in WD\,J1927$-$0355 compared with bulk Earth abundances (relative to magnesium).}
    \label{fig:wdj1927_chi2_contour_Mg}
\end{figure}

The best matching solar system composition for WD\,J1927$-$0355 was bulk Earth. The minimum $\chi^2_{\nu}$ for bulk Earth was 0.8, for $t$\textsubscript{disc}\,$=$\,32\,Myr and $t$\textsubscript{elapse}\,$=$\,5.3\,Myr. The minimum $\chi^2_{\nu}$ for bulk Earth is below 1, implying the error bars on the abundances are somewhat overestimated. The abundances of each element in a parent body accreted 5.3\,Myr ago, with a disc lifetime of 32\,Myr, normalised to magnesium and CI chondrites, are shown in Fig.~\ref{fig:WDJ1927_comparison_to_compositions}. The black horizontal line at zero in Fig.~\ref{fig:WDJ1927_comparison_to_compositions} corresponds to the CI chondrite composition from \citet{Lodders2019}. The abundances of C chondrites are also shown, highlighting that both chondrite and bulk Earth compositions provide a good comparison to the material accreted by WD\,J1927$-$0355. Fig.~\ref{fig:WDJ1927_comparison_to_compositions} is an example of one set of well-matching $t$\textsubscript{disc} and $t$\textsubscript{elapse}, but many other combinations create almost equally good matches because the two free parameters are degenerate, as shown by the $\chi^2_{\nu}$ contour plot in Fig.~\ref{fig:wdj1927_chi2_contour_Mg}. Therefore, the disc lifetime of this system cannot be constrained, however the accretion event is most likely to have occurred between 1\,Myr ago and 10\,Myr ago, depending on the disc lifetime. Table~\ref{tab:min_chi2} shows that C and H chondrites were also good matches to the composition of WD\,J1927$-$0355, but the white dwarf abundances did not match well to achondrites, stony-iron meteorites, or the Earth's crust composition. 

\begin{figure}
    \centering
	\includegraphics[width=\columnwidth]{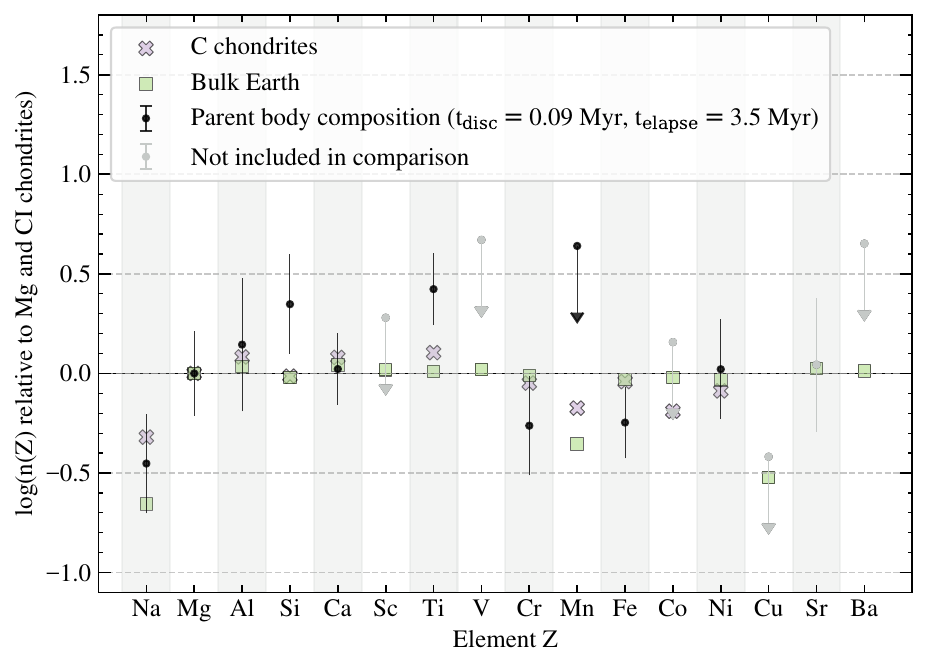}
	\caption{Element abundances for WD\,J2141$-$3300, propagated through the \citet{Jura2009} model for the values of $t$\textsubscript{disc} and $t$\textsubscript{elapse} corresponding to the minimum $\chi^2_{\nu}$, normalised to Mg and CI chondrites, are shown in black. Elements are ordered from left to right in order of increasing atomic weight. Bulk Earth abundances normalised to Mg and CI chondrites are shown by square symbols \citep{Allegre2001}. C chondrite abundances normalised to Mg and CI chondrites are shown by cross symbols \citep{Nittler2004}. Upper limits are denoted with a downwards arrow. CI chondrite abundances are from \citet{Lodders2019}. Error bars are propagated using the errors on abundances from Table~\ref{tab:metal_abundances}. Faded (grey) points represent elements that were not considered in the comparison.}
    \label{fig:WDJ2141_comparison_to_compositions}
\end{figure}

\begin{figure}
    \centering
	\includegraphics[width=\columnwidth]{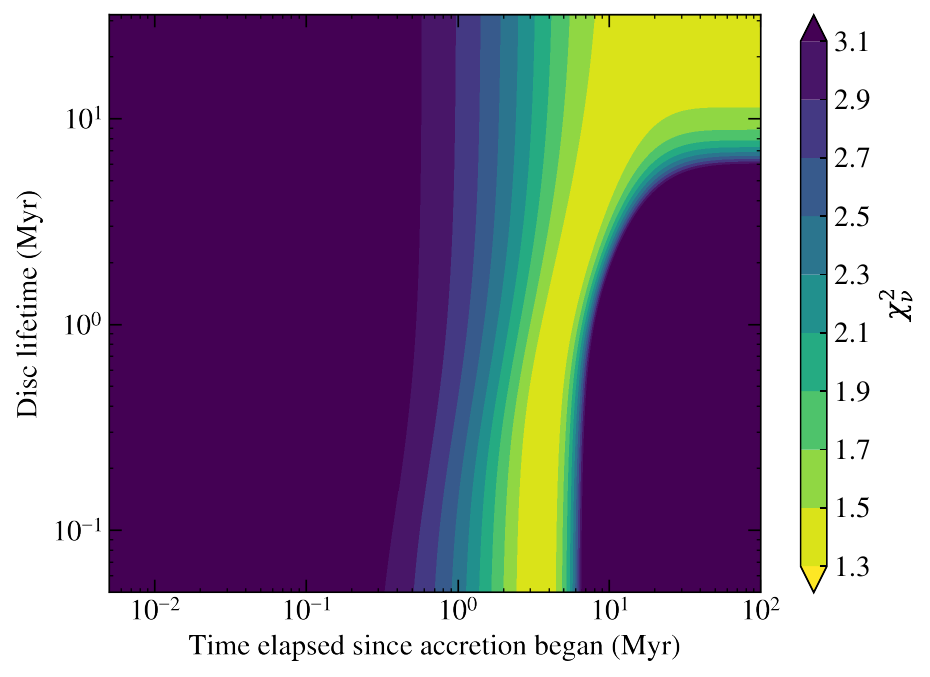}
	\caption{$\chi^2_{\nu}$ contour plot for the abundances in WD\,J2141$-$3300 compared with C chondrite abundances (relative to magnesium).}
    \label{fig:wdj2141_chi2_contour_Mg}
\end{figure}

The best-matching meteorite compositions to WD\,J2141$-$3300 were C chondrites. The minimum $\chi^2_{\nu}$ for C chondrites was 1.3, for $t$\textsubscript{disc}\,$=$\,0.09\,Myr and $t$\textsubscript{elapse}\,$=$\,3.5\,Myr. The corresponding compositions for this combination of $t$\textsubscript{disc} and $t$\textsubscript{elapse} are shown in Fig.~\ref{fig:WDJ2141_comparison_to_compositions}. The inferred titanium and silicon abundances are slightly enhanced compared to C chondrites but remain within 2\,$\sigma$, suggesting no strong deviation from a C chondrite composition. Both the titanium and silicon ratios could be explained by variations of a similar order of magnitude seen in nearby stars. As with WD\,J1927$-$0355, the two free parameters $t$\textsubscript{disc} and $t$\textsubscript{elapse} are degenerate, as shown by the $\chi^2_{\nu}$ contour plot in Fig.~\ref{fig:wdj2141_chi2_contour_Mg}. However, the minimum $\chi^2_{\nu}$ values constrain the time since accretion began to be between 1\,Myr and 6\,Myr ago for shorter disc lifetimes. For the longest disc lifetimes $t$\textsubscript{elapse} is more likely to be greater than 10\,Myr. Fig.~\ref{fig:WDJ2141_comparison_to_compositions} demonstrates that the bulk Earth abundances compare well to the WD\,J2141$-$3300 abundances. Table~\ref{tab:min_chi2} also shows that L chondrites, LL chondrites, and mesosiderites were also good matches to the composition of WD\,J2141$-$3300. The mesosiderite composition is very different to bulk Earth and chondrites, with a iron and nickel-rich composition, and it is only a good comparison for a narrow region of parameter space: at elapsed times greater than 12\,Myr for 4\,--\,6\,Myr disc lifetimes. Achondrites, the Earth's crust composition, and the stony-iron pallasites were poor matches.

\section{Discussion}
\label{sec:discussion_and_interpretation} 

\subsection{Interpretation of WDJ1927-0355 results} 

We determined an estimate of the mass of the parent body at a given $t$\textsubscript{disc} and $t$\textsubscript{elapse} by calculating the mass of all detected elements in the parent body using equation~\ref{eq:m_pb}, plus an assumed mass of oxygen in the parent body. Oxygen is a major component of rock-forming elements, yet is challenging to detect in optical spectra. There are no detections of oxygen absorption lines in either white dwarf, so we adopted an assumed oxygen abundance of log(O/He)\,$=$\,$-$6.55 for both white dwarfs, based on the amount required to achieve a stoichiometric balance with detected rock-forming elements, assuming they were present in the parent body as the following rocky oxides: NaO\textsubscript{2}, MgO, Al\textsubscript{2}O\textsubscript{3}, SiO\textsubscript{2}, CaO, TiO\textsubscript{2}, Cr\textsubscript{2}O\textsubscript{3}, FeO, and NiO. The total mass of heavy elements detected in the convection zone is 6.15$\times$\,10$^{21}$\,g, and including the assumed oxygen abundance this mass goes up to 2.13$\times$\,10$^{22}$\,g. This mass sets a lower limit on the mass of the parent body, and therefore the inferred parent body mass, including the assumed amount of oxygen, ranges from 10$^{22}$\,$-$\,10$^{24}$\,g. The uncertainty in this measurement is dominated by the uncertainty in $t$\textsubscript{elapse}. This mass range spans from roughly the mass of a small moon such as Proteus, a moon of Neptune, up to the mass of a dwarf planet such as Ceres. 

\begin{figure}
    \centering
	\includegraphics[width=\columnwidth]{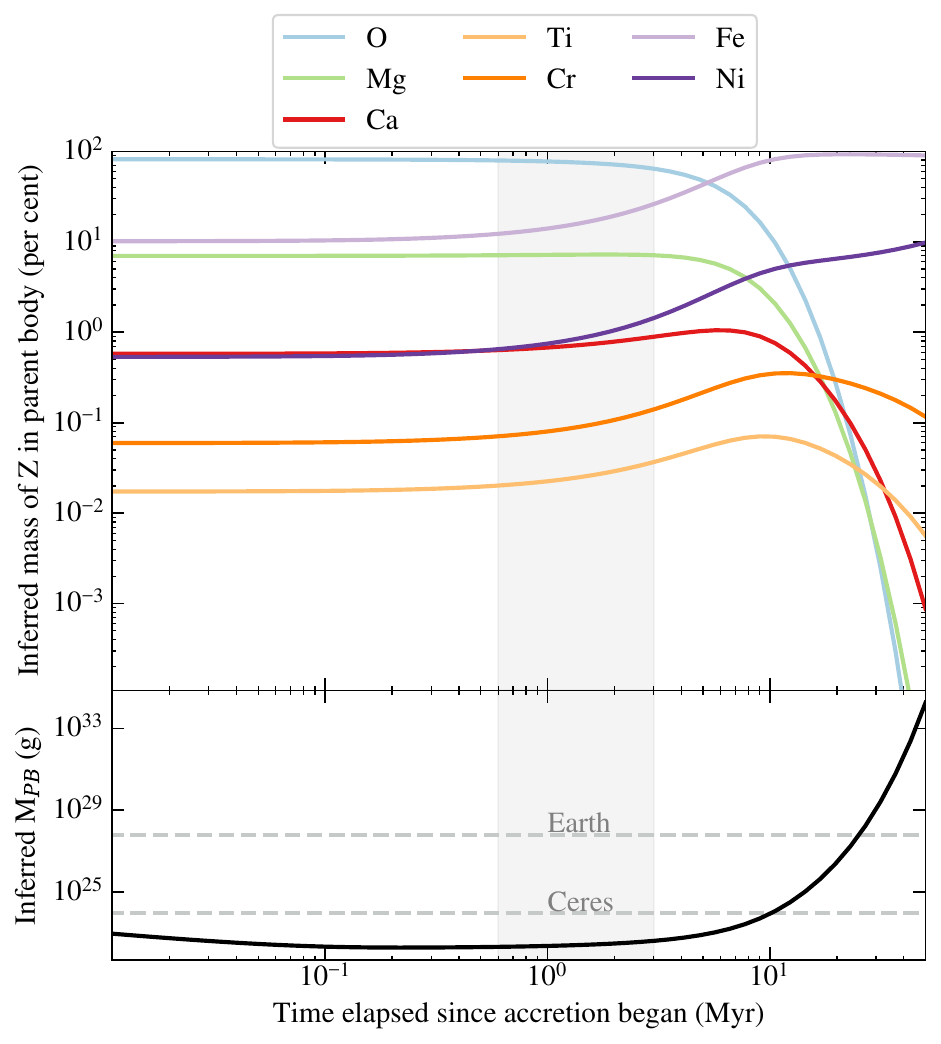}
	\caption{\textbf{Top}: The percentage by mass of all observed elements accreted by WD\,J1927$-$0355, as well as oxygen, as a function of the time since the accretion event began. The disc lifetime is assumed to be 0.05\,Myr. At $t\textsubscript{elapse} = 0$, the order of elements on the plot from top to bottom is: O, Fe, Mg, Ca, Ni, Cr, Ti. \textbf{Bottom}: The total parent body mass (M\textsubscript{PB}) function of the time since the accretion event began. The grey horizontal dashed lines indicate the mass of the dwarf planet Ceres, and the Earth. The grey vertical shaded region is the most likely elapsed time for WD\,J1927$-$0355 since accretion began.}
    \label{fig:wdj1927_mass_percent_1e6}
\end{figure}

Fig.~\ref{fig:wdj1927_mass_percent_1e6} shows the mass ratio profile in the parent body accreted by WD\,J1927$-$0355, as a function of $t$\textsubscript{elapse}, with an illustrative disc lifetime of 0.05\,Myr. Fig.~\ref{fig:wdj1927_mass_percent_1e6} was normalised to 100\,per\,cent for the parent body mass as a function of time since the accretion event began. However, the initial parent body mass required to explain the present day abundances varies depending on the choice of $t$\textsubscript{elapse}, as shown in the lower panel. The shaded region indicates the range of $t$\textsubscript{elapse} that best matches with the bulk Earth.

Given the lack of volatile detections, it is challenging to determine if the parent body is volatile-rich or not, or if the composition is more like bulk Earth or chondrites. Sodium can be considered as a semi-volatile element, however only an upper limit is available. This upper limit indicates that the accreted material is unlikely to be solar or stellar in origin. For non-volatile elements, chondritic and bulk Earth element ratios are very similar, but chondrites have much higher levels of volatiles, hence why some chondrites provide a good match to the accreted material as well as bulk Earth. The sodium upper limit provides a good comparison both to bulk Earth and C chondrites, and therefore both solutions are equally reliable. If the parent body were differentiated, the white dwarf would have still accreted bulk planetary material as opposed to parts of the crust or core of the object, given the best-matching compositions. 

We conclude that WD\,J1927$-$0355 is likely to have accreted bulk material from a rocky planetary body which is the mass of a small moon or dwarf planet. The composition of the parent body is equally likely to be bulk Earth or chondritic. Our conclusion that this system begun accreting material Myrs ago is broadly consistent with many cool DZ systems analysed by \citet{Harrison2021}.

\subsection{Interpretation of WDJ2141-3300 results} 

The best-matching composition to the material accreted by WD\,J2141$-$3300 was C chondrites, but L and LL chondrites were also good matches, as well as bulk Earth (Table~\ref{tab:min_chi2}). The mass of heavy elements in the convection zone is 1.23$\times$\,10$^{22}$\,g, excluding the oxygen assumption, and reaches 3.02$\times$\,10$^{22}$\,g including it. The parent body mass, assuming that the only elements present are all the detected elements and upper limits plus oxygen from rocky oxides, ranges from 10$^{22}$\,$-$\,10$^{24}$\,g. As with WD\,J1927$-$0355, the parent body mass for WD\,J2141$-$3300 corresponds to a small moon or dwarf planet. 

\begin{figure}
    \centering
	\includegraphics[width=\columnwidth]{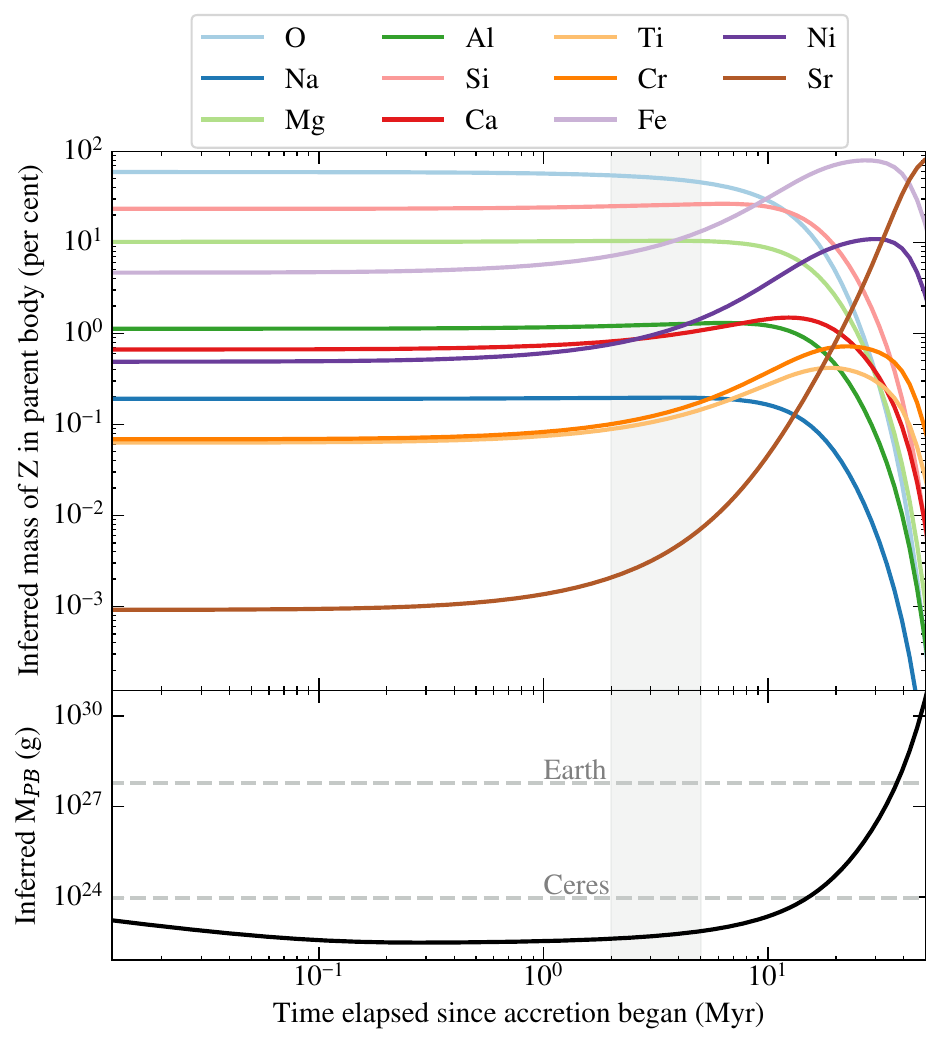}
	\caption{\textbf{Top}: The percentage by mass of all observed elements accreted by WD\,J2141$-$3300, as well as oxygen, as a function of the time since the accretion event began. The disc lifetime is assumed to be 0.05\,Myr. At $t\textsubscript{elapse} = 0$, the order of elements on the plot from top to bottom is: O, Si, Mg, Fe, Al, Ca, Ni, Na, Cr, Ti, Sr. \textbf{Bottom}: The total parent body mass (M\textsubscript{PB}) function of the time since the accretion event began. The grey horizontal dashed lines indicate the mass of the dwarf planet Ceres, and the Earth. The grey vertical shaded region is the most likely elapsed time for WD\,J2141$-$3300 since accretion began.}
    \label{fig:wdj2141_mass_percent_1e6}
\end{figure}

Fig.~\ref{fig:wdj2141_mass_percent_1e6} shows the mass ratio profile in the parent body accreted by WD\,J2141$-$3300, as a function of $t$\textsubscript{elapse}, normalised to 100\,per\,cent for the parent body mass as a function of time since the accretion event began, with an assumed oxygen abundance of log(O/He)\,$=$\,$-$6.55, and a disc lifetime of 0.05\,Myr. The shaded region indicates the range of $t$\textsubscript{elapse} that best matches with the C chondrites, corresponding to the region where the sinking of the material is beginning to dominate over the disc accretion.  

Similarly to WD\,J1927$-$0355, the lack of volatile detections means we cannot reliably determine if the accreted material is rich in volatile elements. There is a detection of the semi-volatile element sodium, which shows that WD\,J2141$-$3300 is not accreting solar or stellar material, and matches with the C chondrite composition as well as the bulk Earth composition, within errors, as shown in Fig.~\ref{fig:WDJ2141_comparison_to_compositions}. More volatile detections are required for a robust conclusion to be drawn, however cool DZ white dwarfs show extreme line blanketing in the blue end of the optical (see Fig.~\ref{fig:both_SED}), and therefore produce minimal flux in the ultraviolet, making the detection of other volatiles challenging. Therefore, we conclude that WD\,J2141$-$3300 has accreted bulk planetary material with a rocky composition, which could be bulk Earth or chondritic in composition. 

\subsection{Comparison with constant accretion rate model}
\label{sec:steady_state}

Throughout this work, we have been using the \citet{Jura2009} model, in which the accretion rate decays exponentially with time after the start of the accretion event. However, a much more commonly used disc model in white dwarf pollution studies assumes instead that the accretion rate is constant until the disc is depleted, after which point it becomes zero \citep{Dupuis1993,Koester2009}. In this model, the accretion of planetary material onto white dwarfs is separated into three somewhat distinct phases: the increasing phase where material is building up onto the surface, the steady state where material is sinking out of the convection zone at roughly the same rate as it is being accreted, and the declining phase where the disc has been depleted and material is sinking out of the atmosphere. The accretion rate is constant in the increasing and steady state phases, and is zero in the declining phase. 

\begin{figure}
    \centering
	\includegraphics[width=\columnwidth]{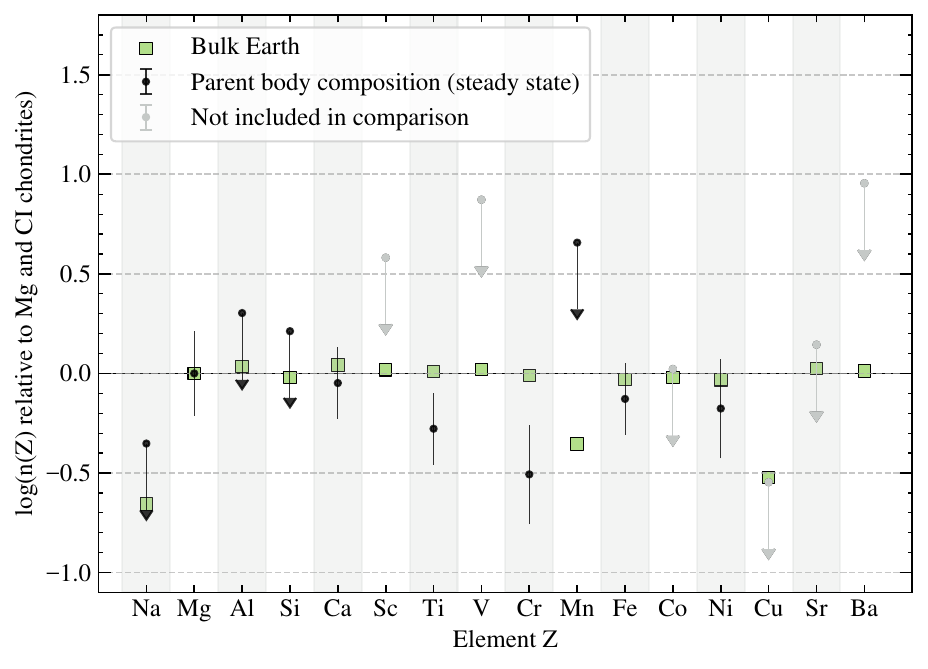}
	\caption{Element abundances for WD\,J1927$-$0355, propagated through the constant accretion rate model for steady state accretion, normalised to Mg and CI chondrites, are shown in black. Elements are ordered from left to right in order of increasing atomic weight. Bulk Earth abundances normalised to Mg and CI chondrites are shown by square symbols \citep{Nittler2004}. Upper limits are denoted with a downwards arrow. Error bars are propagated using the errors on abundances from Table~\ref{tab:metal_abundances}. CI chondrite abundances are from \citet{Lodders2019}. Faded (grey) points represent elements that were not considered in the comparison.}
    \label{fig:wdj1927_const_acc_model}
\end{figure}

\begin{figure}
    \centering
	\includegraphics[width=\columnwidth]{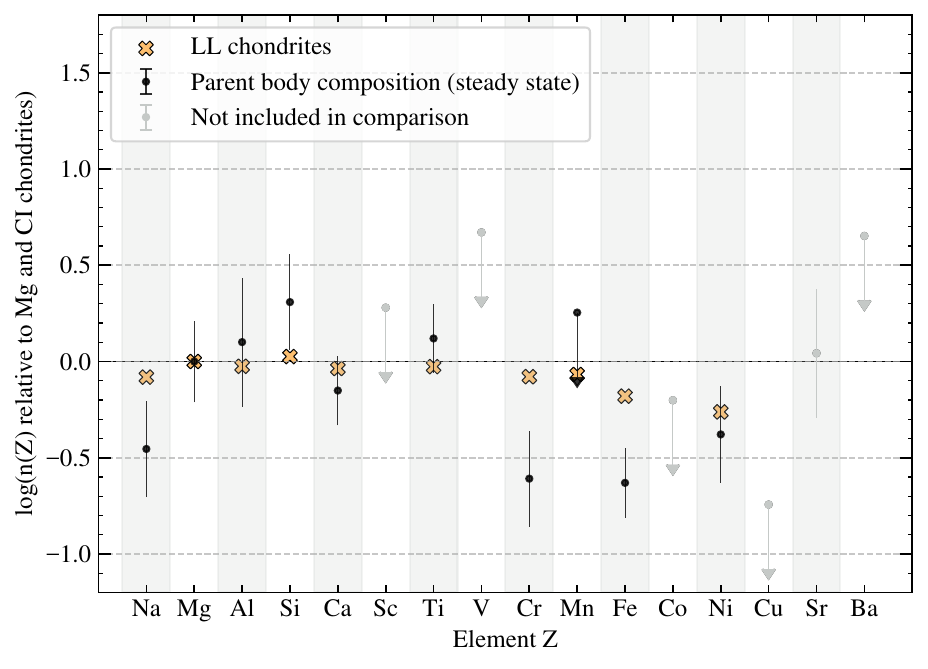}
	\caption{Element abundances for WD\,J2141$-$3300, propagated through the constant accretion rate model for steady state accretion, normalised to Mg and CI chondrites, are shown in black. Elements are ordered from left to right in order of increasing atomic weight. LL chondrite abundances normalised to Mg and CI chondrites are shown in by cross symbols \citep{Nittler2004}. Upper limits are denoted with a downwards arrow. Error bars are propagated using the errors on abundances from Table~\ref{tab:metal_abundances}. CI chondrite abundances are from \citet{Lodders2019}. Faded (grey) points represent elements that were not considered in the comparison.}
    \label{fig:wdj2141_const_acc_model}
\end{figure}

We calculated the minimised $\chi^2_{\nu}$ for the constant accretion rate model, comparing the WD\,J1927$-$0355 abundances to the solar system compositions in Table~\ref{tab:min_chi2}. To calculate $\chi^2_{\nu}$, we used a similar method to that outlined in Section~\ref{sec:methods}, but with equations from \citet{Koester2009} to model the constant accretion rate. The best-matching composition was again bulk Earth, with a minimised $\chi^2_{\nu}$ of 2.5 which corresponded to steady state accretion. The abundances of the individual elements used in this calculation compared to bulk Earth are shown in Fig.~\ref{fig:wdj1927_const_acc_model}. The best-matching composition under the constant accretion rate model for WD\,J2141$-$3300 abundances was LL chondrites, with the minimum $\chi^2_{\nu}$ being 2.4, also when the system was in a steady state. The abundances of the individual elements used in this calculation compared to LL chondrites are shown in Fig.~\ref{fig:wdj2141_const_acc_model}. These results are independent of the disc lifetime, since theoretically the disc lifetime could have any duration if both white dwarfs are accreting in the steady state.

Figures~\ref{fig:wdj1927_const_acc_model} and \ref{fig:wdj2141_const_acc_model} highlight that the error bars on the abundances are sufficiently large that the choice of disc model, constant accretion rate or exponentially decaying accretion rate, does not affect the best-matching compositions. Both white dwarfs were best matched by a steady state accretion scenario under the constant accretion rate model. Metals remain in the atmosphere of cool helium-atmosphere white dwarfs for Myrs, and therefore steady state accretion requires a sustained source of accretion over a period much longer than the settling timescales. This could be possible if many smaller parent bodies are being consistently accreted onto the white dwarf over long timescales, therefore replenishing the disc. However, the exponentially decaying model assumes that the disc is not replenished, and provides a solution of which there is no analogue in the constant accretion rate model: that the metals are sinking slightly faster than the rate at which they are being accreted, but accretion has not ceased entirely. 

\subsection{Future work}
\label{sec:future}

The results of applying the exponentially decaying disc model to WD\,J1927$-$0355 and WD\,J2141$-$3300 demonstrate the importance of considering disc lifetimes in the analysis of polluted white dwarfs. To date, 62 helium-atmosphere white dwarfs have been observed to contain six or more metals in their atmosphere originating from the accretion of planetary debris \citep{Hollands2017,Williams2024}. Applying the \citet{Jura2009} disc model to this large group of white dwarfs would enable further constraints on the masses and types of parent bodies providing the pollution.

The exponentially decaying disc model is not necessarily a physical improvement on the constant accretion rate model, as there is no way to know without a direct disc detection whether the disc is being replenished or not. Both stars should be followed up with infrared facilities such as \textit{JWST}, which has the capability to detect faint white dwarf dust discs \citep{Farihi2025}, and can further constrain the phase of accretion of these systems. The exponentially decaying disc model assumes that no further material is being added to the disc, whereas a model assuming steady state accretion could imply that material is continuously being added to the disc, similar to the scenarios described in e.g. \citet{Wyatt2014}. However, the exponentially decaying disc model provides a wider range of parameter space in order to compare observations with solar system compositions, and a scenario in which material is decaying slightly faster than it is being accreted matches better with both white dwarf observations than any part of the constant accretion model. 

A larger sample of observations of white dwarf debris discs are required in order to reduce the uncertainty on the characteristic disc lifetime. A set of observations of cool helium-atmosphere white dwarfs with many photospheric metals should additionally be analysed using the framework outlined in this work, in order to begin distinguishing if the exponentially decaying disc model makes a systematic difference on the analysis, and whether it should be used going forward in the analysis of these systems.

\section{Conclusions}

An exponentially decaying disc model from \citet{Jura2009} was adopted in the analysis of the planetary material accreted by two cool helium-atmosphere white dwarfs: WD\,J1927$-$0355 and WD\,J2141$-$3300. These stars have detections of six and ten metals in their atmosphere respectively. Optical spectra of these two white dwarfs, taken from the HIRES instrument on the Keck telescope, were fitted with tailored white dwarf atmosphere models incorporating detected elements plus upper limits \citep{Koester2010}. The white dwarf atmosphere models struggled to fit the narrow H$\alpha$ features in both spectra. Similarly narrow H$\alpha$ features have been observed in other cool, polluted helium-atmosphere white dwarfs, and further improvements to model atmosphere codes are needed to obtain accurate H/He ratios. The \citet{Jura2009} model was applied to the abundances from the spectroscopic fits in order to constrain the type of accreted material and the phase of accretion.

The \citet{Jura2009} model considers a disc for which accretion rates decay exponentially, which is physically motivated by derivations from \citet{Lodato2008} for a disc in which viscosity is proportional to radius. The time since the accretion event began and the characteristic lifetime of the disc were set as free parameters, and $\chi^2_{\nu}$ contour plots constrained the mass and type of the accreted material. This approach considers disc lifetime as a free parameter in the \citet{Jura2009} disc model in order to constrain the composition of the material accreted by the white dwarfs, and motivates further observations and theory in order to constrain white dwarf disc lifetimes.

It was found that WD\,J1927$-$0355 and WD\,J2141$-$3300 are likely to be experiencing accretion of material similar to bulk Earth or C chondrites in composition. The only available element to distinguish between these compositions is sodium, and in both cases the sodium detection or upper limit matches with both bulk Earth and chondrites within errors. These compositions are typical for white dwarf planetary systems \citep{Hollands2018_dz}. The best matches to the observations of both stars implied that they are being observed Myrs into an accretion event, and their discs must therefore be slightly depleted. The masses of the parent bodies accreted onto both systems span a range from 10$^{22}$\,$-$10$^{24}$\,g, which means both systems are likely to have accreted bulk planetary material roughly the mass of a large asteroid or dwarf planet. The two-dimensional parameter space of disc lifetime and time since the accretion event began provided constraints on both the phase of accretion and the total mass of the accreted material, but could not constrain the duration of the accretion event. As a comparison, the constant accretion rate model places both of these white dwarfs in the steady state phase.

These two white dwarfs add to the growing number of polluted white dwarfs that have been analysed in detail, and they provide further evidence that typical white dwarf planetary systems contain rocky planetesimals with composition ratios similar to bulk Earth and chondritic compositions. A larger sample of cool white dwarfs with multiple photospheric metals should be analysed with this same method in the future, in order to identify targets for infrared follow-up. Additionally, more robust measurements of disc lifetimes would help to further constrain the masses of the accreted parent bodies. 

\section*{Acknowledgements}
We thank the referee for their helpful comments and suggestions for improving this manuscript. MOB and PET received funding from the European Research Council under the European Union’s Horizon 2020 research and innovation programme number 101002408. MOB would like to thank Rebecca Nealon, Andrew Swan, Isabella Trierweiler, and Jamie Williams for helpful discussions regarding the interpretation of the results presented in this work. We thank Ben Zuckerman for his assistance in obtaining the HIRES spectra.

Some of the data presented herein were obtained at the W. M. Keck Observatory, which is operated as a scientific partnership among the California Institute of Technology, the University of California and the National Aeronautics and Space Administration. The Observatory was made possible by the generous financial support of the W. M. Keck Foundation. The authors wish to recognise and acknowledge the very significant cultural role and reverence that the summit of Maunakea has always had within the indigenous Hawaiian community. We are most fortunate to have the opportunity to conduct observations from this mountain. This work has made use of data from the European Space Agency (ESA) mission \textit{Gaia} (\url{https://www.cosmos.esa.int/gaia}), processed by the \textit{Gaia} Data Processing and Analysis Consortium (DPAC, \url{https://www.cosmos.esa.int/web/gaia/dpac/consortium}). Funding for the DPAC has been provided by national institutions, in particular the institutions participating in the \textit{Gaia} Multilateral Agreement. 

This research has made use of NASA’s Astrophysics Data System; the SIMBAD database, operated at CDS, Strasbourg, France; and the VizieR service. This work made use of \texttt{Astropy}:\footnote{http://www.astropy.org} a community-developed core \textsc{Python} package and an ecosystem of tools and resources for astronomy \citep{astropy:2013, astropy:2018, astropy:2022}. This work also made use of the \textsc{Python} package \texttt{scipy} \citep{2020SciPy-NMeth}. This work made use of \textsc{Python} scripts for reading echelle spectra files, created by \citet{Gullikson2015}. 

\section*{Data Availability}
The reduced spectra presented in this article will be shared on reasonable request to the corresponding author.



\bibliographystyle{mnras}
\bibliography{mybib} 

\begin{thebibliography}{}
\makeatletter
\relax
\def\mn@urlcharsother{\let\do\@makeother \do\$\do\&\do\#\do\^\do\_\do\%\do\~}
\def\mn@doi{\begingroup\mn@urlcharsother \@ifnextchar [ {\mn@doi@} {\mn@doi@[]}}
\def\mn@doi@[#1]#2{\def\@tempa{#1}\ifx\@tempa\@empty \href {http://dx.doi.org/#2} {doi:#2}\else \href {http://dx.doi.org/#2} {#1}\fi \endgroup}
\def\mn@eprint#1#2{\mn@eprint@#1:#2::\@nil}
\def\mn@eprint@arXiv#1{\href {http://arxiv.org/abs/#1} {{\tt arXiv:#1}}}
\def\mn@eprint@dblp#1{\href {http://dblp.uni-trier.de/rec/bibtex/#1.xml} {dblp:#1}}
\def\mn@eprint@#1:#2:#3:#4\@nil{\def\@tempa {#1}\def\@tempb {#2}\def\@tempc {#3}\ifx \@tempc \@empty \let \@tempc \@tempb \let \@tempb \@tempa \fi \ifx \@tempb \@empty \def\@tempb {arXiv}\fi \@ifundefined {mn@eprint@\@tempb}{\@tempb:\@tempc}{\expandafter \expandafter \csname mn@eprint@\@tempb\endcsname \expandafter{\@tempc}}}

\bibitem[\protect\citeauthoryear{{All{\`e}gre}, {Manh{\`e}s}  \& {Lewin}}{{All{\`e}gre} et~al.}{2001}]{Allegre2001}
{All{\`e}gre} C.,  {Manh{\`e}s} G.,   {Lewin} {\'E}.,  2001, \mn@doi [Earth and Planetary Science Letters] {10.1016/S0012-821X(00)00359-9}, \href {https://ui.adsabs.harvard.edu/abs/2001E&PSL.185...49A} {185, 49}

\bibitem[\protect\citeauthoryear{{Astropy Collaboration} et~al.,}{{Astropy Collaboration} et~al.}{2013}]{astropy:2013}
{Astropy Collaboration} et~al., 2013, \mn@doi [\aap] {10.1051/0004-6361/201322068}, \href {http://adsabs.harvard.edu/abs/2013A%26A...558A..33A} {558, A33}

\bibitem[\protect\citeauthoryear{{Astropy Collaboration} et~al.,}{{Astropy Collaboration} et~al.}{2018}]{astropy:2018}
{Astropy Collaboration} et~al., 2018, \mn@doi [\aj] {10.3847/1538-3881/aabc4f}, \href {https://ui.adsabs.harvard.edu/abs/2018AJ....156..123A} {156, 123}

\bibitem[\protect\citeauthoryear{{Astropy Collaboration} et~al.,}{{Astropy Collaboration} et~al.}{2022}]{astropy:2022}
{Astropy Collaboration} et~al., 2022, \mn@doi [\apj] {10.3847/1538-4357/ac7c74}, \href {https://ui.adsabs.harvard.edu/abs/2022ApJ...935..167A} {935, 167}

\bibitem[\protect\citeauthoryear{{Bagnulo} \& {Landstreet}}{{Bagnulo} \& {Landstreet}}{2019}]{Bagnulo2019}
{Bagnulo} S.,  {Landstreet} J.~D.,  2019, \mn@doi [\aap] {10.1051/0004-6361/201936068}, \href {https://ui.adsabs.harvard.edu/abs/2019A&A...630A..65B} {630, A65}

\bibitem[\protect\citeauthoryear{{Bagnulo}, {Landstreet}, {Farihi}, {Folsom}, {Hollands}  \& {Fossati}}{{Bagnulo} et~al.}{2024a}]{Bagnulo2024b}
{Bagnulo} S.,  {Landstreet} J.~D.,  {Farihi} J.,  {Folsom} C.~P.,  {Hollands} M.~A.,   {Fossati} L.,  2024a, \mn@doi [\aap] {10.1051/0004-6361/202451215}, \href {https://ui.adsabs.harvard.edu/abs/2024A&A...688L..14B} {688, L14}

\bibitem[\protect\citeauthoryear{{Bagnulo}, {Farihi}, {Landstreet}  \& {Folsom}}{{Bagnulo} et~al.}{2024b}]{Bagnulo2024a}
{Bagnulo} S.,  {Farihi} J.,  {Landstreet} J.~D.,   {Folsom} C.~P.,  2024b, \mn@doi [\apjl] {10.3847/2041-8213/ad2619}, \href {https://ui.adsabs.harvard.edu/abs/2024ApJ...963L..22B} {963, L22}

\bibitem[\protect\citeauthoryear{{Brouwers}, {Bonsor}  \& {Malamud}}{{Brouwers} et~al.}{2022}]{Brouwers2022}
{Brouwers} M.~G.,  {Bonsor} A.,   {Malamud} U.,  2022, \mn@doi [\mnras] {10.1093/mnras/stab3009}, \href {https://ui.adsabs.harvard.edu/abs/2022MNRAS.509.2404B} {509, 2404}

\bibitem[\protect\citeauthoryear{{Buchan}, {Bonsor}, {Shorttle}, {Wade}, {Harrison}, {Noack}  \& {Koester}}{{Buchan} et~al.}{2022}]{Buchan2022}
{Buchan} A.~M.,  {Bonsor} A.,  {Shorttle} O.,  {Wade} J.,  {Harrison} J.,  {Noack} L.,   {Koester} D.,  2022, \mn@doi [\mnras] {10.1093/mnras/stab3624}, \href {https://ui.adsabs.harvard.edu/abs/2022MNRAS.510.3512B} {510, 3512}

\bibitem[\protect\citeauthoryear{{Chambers} et~al.,}{{Chambers} et~al.}{2016}]{PanSTARRS2016}
{Chambers} K.~C.,  et~al., 2016, arXiv e-prints, \href {https://ui.adsabs.harvard.edu/abs/2016arXiv161205560C} {p. arXiv:1612.05560}

\bibitem[\protect\citeauthoryear{{Coutu}, {Dufour}, {Bergeron}, {Blouin}, {Loranger}, {Allard}  \& {Dunlap}}{{Coutu} et~al.}{2019}]{Coutu2019}
{Coutu} S.,  {Dufour} P.,  {Bergeron} P.,  {Blouin} S.,  {Loranger} E.,  {Allard} N.~F.,   {Dunlap} B.~H.,  2019, \mn@doi [\apj] {10.3847/1538-4357/ab46b9}, \href {https://ui.adsabs.harvard.edu/abs/2019ApJ...885...74C} {885, 74}

\bibitem[\protect\citeauthoryear{{Cunningham}, {Tremblay}, {Freytag}, {Ludwig}  \& {Koester}}{{Cunningham} et~al.}{2019}]{Cunningham2019}
{Cunningham} T.,  {Tremblay} P.-E.,  {Freytag} B.,  {Ludwig} H.-G.,   {Koester} D.,  2019, \mn@doi [\mnras] {10.1093/mnras/stz1759}, \href {https://ui.adsabs.harvard.edu/abs/2019MNRAS.488.2503C} {488, 2503}

\bibitem[\protect\citeauthoryear{{Cunningham} et~al.,}{{Cunningham} et~al.}{2021}]{Cunningham2021}
{Cunningham} T.,  et~al., 2021, \mn@doi [\mnras] {10.1093/mnras/stab553}, \href {https://ui.adsabs.harvard.edu/abs/2021MNRAS.503.1646C} {503, 1646}

\bibitem[\protect\citeauthoryear{{Debes} \& {Sigurdsson}}{{Debes} \& {Sigurdsson}}{2002}]{Debes2002}
{Debes} J.~H.,  {Sigurdsson} S.,  2002, \mn@doi [\apj] {10.1086/340291}, \href {https://ui.adsabs.harvard.edu/abs/2002ApJ...572..556D} {572, 556}

\bibitem[\protect\citeauthoryear{{Dennihy} et~al.,}{{Dennihy} et~al.}{2020}]{Dennihy2020}
{Dennihy} E.,  et~al., 2020, \mn@doi [\apj] {10.3847/1538-4357/abc339}, \href {https://ui.adsabs.harvard.edu/abs/2020ApJ...905....5D} {905, 5}

\bibitem[\protect\citeauthoryear{{Doyle}, {Klein}, {Schlichting}  \& {Young}}{{Doyle} et~al.}{2020}]{Doyle2020}
{Doyle} A.~E.,  {Klein} B.,  {Schlichting} H.~E.,   {Young} E.~D.,  2020, \mn@doi [\apj] {10.3847/1538-4357/abad9a}, \href {https://ui.adsabs.harvard.edu/abs/2020ApJ...901...10D} {901, 10}

\bibitem[\protect\citeauthoryear{{Doyle}, {Desch}  \& {Young}}{{Doyle} et~al.}{2021}]{Doyle2021}
{Doyle} A.~E.,  {Desch} S.~J.,   {Young} E.~D.,  2021, \mn@doi [\apjl] {10.3847/2041-8213/abd9ba}, \href {https://ui.adsabs.harvard.edu/abs/2021ApJ...907L..35D} {907, L35}

\bibitem[\protect\citeauthoryear{{Doyle} et~al.,}{{Doyle} et~al.}{2023}]{Doyle2023}
{Doyle} A.~E.,  et~al., 2023, \mn@doi [\apj] {10.3847/1538-4357/acbd44}, \href {https://ui.adsabs.harvard.edu/abs/2023ApJ...950...93D} {950, 93}

\bibitem[\protect\citeauthoryear{{Dufour} et~al.,}{{Dufour} et~al.}{2007}]{Dufour2007_dz}
{Dufour} P.,  et~al., 2007, \mn@doi [\apj] {10.1086/518468}, \href {https://ui.adsabs.harvard.edu/abs/2007ApJ...663.1291D} {663, 1291}

\bibitem[\protect\citeauthoryear{{Dupuis}, {Fontaine}, {Pelletier}  \& {Wesemael}}{{Dupuis} et~al.}{1993}]{Dupuis1993}
{Dupuis} J.,  {Fontaine} G.,  {Pelletier} C.,   {Wesemael} F.,  1993, \mn@doi [\apjs] {10.1086/191746}, \href {https://ui.adsabs.harvard.edu/abs/1993ApJS...84...73D} {84, 73}

\bibitem[\protect\citeauthoryear{{Elms}, {Tremblay}, {G{\"a}nsicke}, {Koester}, {Hollands}, {Gentile Fusillo}, {Cunningham}  \& {Apps}}{{Elms} et~al.}{2022}]{Elms2022}
{Elms} A.~K.,  {Tremblay} P.-E.,  {G{\"a}nsicke} B.~T.,  {Koester} D.,  {Hollands} M.~A.,  {Gentile Fusillo} N.~P.,  {Cunningham} T.,   {Apps} K.,  2022, \mn@doi [\mnras] {10.1093/mnras/stac2908}, \href {https://ui.adsabs.harvard.edu/abs/2022MNRAS.517.4557E} {517, 4557}

\bibitem[\protect\citeauthoryear{{Farihi}}{{Farihi}}{2016}]{Farihi2016}
{Farihi} J.,  2016, \mn@doi [\nar] {10.1016/j.newar.2016.03.001}, \href {https://ui.adsabs.harvard.edu/abs/2016NewAR..71....9F} {71, 9}

\bibitem[\protect\citeauthoryear{{Farihi}, {Jura}  \& {Zuckerman}}{{Farihi} et~al.}{2009}]{Farihi2009}
{Farihi} J.,  {Jura} M.,   {Zuckerman} B.,  2009, \mn@doi [\apj] {10.1088/0004-637X/694/2/805}, \href {https://ui.adsabs.harvard.edu/abs/2009ApJ...694..805F} {694, 805}

\bibitem[\protect\citeauthoryear{{Farihi}, {Bond}, {Dufour}, {Haghighipour}, {Schaefer}, {Holberg}, {Barstow}  \& {Burleigh}}{{Farihi} et~al.}{2013}]{Farihi2013}
{Farihi} J.,  {Bond} H.~E.,  {Dufour} P.,  {Haghighipour} N.,  {Schaefer} G.~H.,  {Holberg} J.~B.,  {Barstow} M.~A.,   {Burleigh} M.~R.,  2013, \mn@doi [\mnras] {10.1093/mnras/sts677}, \href {https://ui.adsabs.harvard.edu/abs/2013MNRAS.430..652F} {430, 652}

\bibitem[\protect\citeauthoryear{{Farihi}, {Robert}  \& {Walters}}{{Farihi} et~al.}{2024}]{Farihi2024}
{Farihi} J.,  {Robert} A.,   {Walters} N.,  2024, \mn@doi [\mnras] {10.1093/mnrasl/slae014}, \href {https://ui.adsabs.harvard.edu/abs/2024MNRAS.529L.164F} {529, L164}

\bibitem[\protect\citeauthoryear{{Farihi}, {Su}, {Melis}, {Kenyon}, {Swan}, {Redfield}, {Wyatt}  \& {Debes}}{{Farihi} et~al.}{2025}]{Farihi2025}
{Farihi} J.,  {Su} K.~Y.~L.,  {Melis} C.,  {Kenyon} S.~J.,  {Swan} A.,  {Redfield} S.,  {Wyatt} M.~C.,   {Debes} J.~H.,  2025, \mn@doi [arXiv e-prints] {10.48550/arXiv.2501.18338}, \href {https://ui.adsabs.harvard.edu/abs/2025arXiv250118338F} {p. arXiv:2501.18338}

\bibitem[\protect\citeauthoryear{{G{\"a}nsicke}, {Marsh}, {Southworth}  \& {Rebassa-Mansergas}}{{G{\"a}nsicke} et~al.}{2006}]{Gaensicke2006}
{G{\"a}nsicke} B.~T.,  {Marsh} T.~R.,  {Southworth} J.,   {Rebassa-Mansergas} A.,  2006, \mn@doi [Science] {10.1126/science.1135033}, \href {https://ui.adsabs.harvard.edu/abs/2006Sci...314.1908G} {314, 1908}

\bibitem[\protect\citeauthoryear{{G{\"a}nsicke}, {Marsh}  \& {Southworth}}{{G{\"a}nsicke} et~al.}{2007}]{Gaensicke2007}
{G{\"a}nsicke} B.~T.,  {Marsh} T.~R.,   {Southworth} J.,  2007, \mn@doi [\mnras] {10.1111/j.1745-3933.2007.00343.x}, \href {https://ui.adsabs.harvard.edu/abs/2007MNRAS.380L..35G} {380, L35}

\bibitem[\protect\citeauthoryear{{G{\"a}nsicke}, {Koester}, {Marsh}, {Rebassa-Mansergas}  \& {Southworth}}{{G{\"a}nsicke} et~al.}{2008}]{Gaensicke2008}
{G{\"a}nsicke} B.~T.,  {Koester} D.,  {Marsh} T.~R.,  {Rebassa-Mansergas} A.,   {Southworth} J.,  2008, \mn@doi [\mnras] {10.1111/j.1745-3933.2008.00565.x}, \href {https://ui.adsabs.harvard.edu/abs/2008MNRAS.391L.103G} {391, L103}

\bibitem[\protect\citeauthoryear{{G{\"a}nsicke}, {Koester}, {Farihi}, {Girven}, {Parsons}  \& {Breedt}}{{G{\"a}nsicke} et~al.}{2012}]{Gaensicke2012}
{G{\"a}nsicke} B.~T.,  {Koester} D.,  {Farihi} J.,  {Girven} J.,  {Parsons} S.~G.,   {Breedt} E.,  2012, \mn@doi [\mnras] {10.1111/j.1365-2966.2012.21201.x}, \href {https://ui.adsabs.harvard.edu/abs/2012MNRAS.424..333G} {424, 333}

\bibitem[\protect\citeauthoryear{{G{\"a}nsicke}, {Koester}, {Farihi}  \& {Toloza}}{{G{\"a}nsicke} et~al.}{2018}]{Gaensicke2018}
{G{\"a}nsicke} B.~T.,  {Koester} D.,  {Farihi} J.,   {Toloza} O.,  2018, \mn@doi [\mnras] {10.1093/mnras/sty2526}, \href {https://ui.adsabs.harvard.edu/abs/2018MNRAS.481.4323G} {481, 4323}

\bibitem[\protect\citeauthoryear{{Gentile Fusillo} et~al.,}{{Gentile Fusillo} et~al.}{2019}]{Gentile2019}
{Gentile Fusillo} N.~P.,  et~al., 2019, \mn@doi [\mnras] {10.1093/mnras/sty3016}, \href {https://ui.adsabs.harvard.edu/abs/2019MNRAS.482.4570G} {482, 4570}

\bibitem[\protect\citeauthoryear{{Gentile Fusillo} et~al.,}{{Gentile Fusillo} et~al.}{2021}]{Gentile2021_gas}
{Gentile Fusillo} N.~P.,  et~al., 2021, \mn@doi [\mnras] {10.1093/mnras/stab992}, \href {https://ui.adsabs.harvard.edu/abs/2021MNRAS.504.2707G} {504, 2707}

\bibitem[\protect\citeauthoryear{{Girven}, {Brinkworth}, {Farihi}, {G{\"a}nsicke}, {Hoard}, {Marsh}  \& {Koester}}{{Girven} et~al.}{2012}]{Girven2012}
{Girven} J.,  {Brinkworth} C.~S.,  {Farihi} J.,  {G{\"a}nsicke} B.~T.,  {Hoard} D.~W.,  {Marsh} T.~R.,   {Koester} D.,  2012, \mn@doi [\apj] {10.1088/0004-637X/749/2/154}, \href {https://ui.adsabs.harvard.edu/abs/2012ApJ...749..154G} {749, 154}

\bibitem[\protect\citeauthoryear{{Guidry} et~al.,}{{Guidry} et~al.}{2021}]{Guidry2021}
{Guidry} J.~A.,  et~al., 2021, \mn@doi [\apj] {10.3847/1538-4357/abee68}, \href {https://ui.adsabs.harvard.edu/abs/2021ApJ...912..125G} {912, 125}

\bibitem[\protect\citeauthoryear{Gullikson}{Gullikson}{2015}]{Gullikson2015}
Gullikson K.,  2015, General-Scripts\_v1.0, \url {https://doi.org/10.5281/zenodo.10013}

\bibitem[\protect\citeauthoryear{{Harrison}, {Bonsor}, {Kama}, {Buchan}, {Blouin}  \& {Koester}}{{Harrison} et~al.}{2021}]{Harrison2021}
{Harrison} J. H.~D.,  {Bonsor} A.,  {Kama} M.,  {Buchan} A.~M.,  {Blouin} S.,   {Koester} D.,  2021, \mn@doi [\mnras] {10.1093/mnras/stab736}, \href {https://ui.adsabs.harvard.edu/abs/2021MNRAS.504.2853H} {504, 2853}

\bibitem[\protect\citeauthoryear{{Hernandez}, {Schreiber}, {Landstreet}, {Bagnulo}, {Parsons}, {Chavarria}, {Toloza}  \& {Bell}}{{Hernandez} et~al.}{2024}]{Hernandez2024}
{Hernandez} M.~S.,  {Schreiber} M.~R.,  {Landstreet} J.~D.,  {Bagnulo} S.,  {Parsons} S.~G.,  {Chavarria} M.,  {Toloza} O.,   {Bell} K.~J.,  2024, \mn@doi [\mnras] {10.1093/mnras/stae307}, \href {https://ui.adsabs.harvard.edu/abs/2024MNRAS.528.6056H} {528, 6056}

\bibitem[\protect\citeauthoryear{{Hollands}, {Koester}, {Alekseev}, {Herbert}  \& {G{\"a}nsicke}}{{Hollands} et~al.}{2017}]{Hollands2017}
{Hollands} M.~A.,  {Koester} D.,  {Alekseev} V.,  {Herbert} E.~L.,   {G{\"a}nsicke} B.~T.,  2017, \mn@doi [\mnras] {10.1093/mnras/stx250}, \href {https://ui.adsabs.harvard.edu/abs/2017MNRAS.467.4970H} {467, 4970}

\bibitem[\protect\citeauthoryear{{Hollands}, {G{\"a}nsicke}  \& {Koester}}{{Hollands} et~al.}{2018}]{Hollands2018_dz}
{Hollands} M.~A.,  {G{\"a}nsicke} B.~T.,   {Koester} D.,  2018, \mn@doi [\mnras] {10.1093/mnras/sty592}, \href {https://ui.adsabs.harvard.edu/abs/2018MNRAS.477...93H} {477, 93}

\bibitem[\protect\citeauthoryear{{Hollands}, {Tremblay}, {G{\"a}nsicke}, {Koester}  \& {Gentile-Fusillo}}{{Hollands} et~al.}{2021}]{Hollands2021}
{Hollands} M.~A.,  {Tremblay} P.-E.,  {G{\"a}nsicke} B.~T.,  {Koester} D.,   {Gentile-Fusillo} N.~P.,  2021, \mn@doi [Nature Astronomy] {10.1038/s41550-020-01296-7}, \href {https://ui.adsabs.harvard.edu/abs/2021NatAs...5..451H} {5, 451}

\bibitem[\protect\citeauthoryear{{Hoskin} et~al.,}{{Hoskin} et~al.}{2020}]{Hoskin2020}
{Hoskin} M.~J.,  et~al., 2020, \mn@doi [\mnras] {10.1093/mnras/staa2717}, \href {https://ui.adsabs.harvard.edu/abs/2020MNRAS.499..171H} {499, 171}

\bibitem[\protect\citeauthoryear{{Jura}}{{Jura}}{2003}]{Jura2003}
{Jura} M.,  2003, \mn@doi [\apjl] {10.1086/374036}, \href {https://ui.adsabs.harvard.edu/abs/2003ApJ...584L..91J} {584, L91}

\bibitem[\protect\citeauthoryear{{Jura} \& {Xu}}{{Jura} \& {Xu}}{2010}]{Jura2010}
{Jura} M.,  {Xu} S.,  2010, \mn@doi [\aj] {10.1088/0004-6256/140/5/1129}, \href {https://ui.adsabs.harvard.edu/abs/2010AJ....140.1129J} {140, 1129}

\bibitem[\protect\citeauthoryear{{Jura}, {Farihi}, {Zuckerman}  \& {Becklin}}{{Jura} et~al.}{2007a}]{Jura2007b}
{Jura} M.,  {Farihi} J.,  {Zuckerman} B.,   {Becklin} E.~E.,  2007a, \mn@doi [\aj] {10.1086/512734}, \href {https://ui.adsabs.harvard.edu/abs/2007AJ....133.1927J} {133, 1927}

\bibitem[\protect\citeauthoryear{{Jura}, {Farihi}  \& {Zuckerman}}{{Jura} et~al.}{2007b}]{Jura2007a}
{Jura} M.,  {Farihi} J.,   {Zuckerman} B.,  2007b, \mn@doi [\apj] {10.1086/518767}, \href {https://ui.adsabs.harvard.edu/abs/2007ApJ...663.1285J} {663, 1285}

\bibitem[\protect\citeauthoryear{{Jura}, {Muno}, {Farihi}  \& {Zuckerman}}{{Jura} et~al.}{2009}]{Jura2009}
{Jura} M.,  {Muno} M.~P.,  {Farihi} J.,   {Zuckerman} B.,  2009, \mn@doi [\apj] {10.1088/0004-637X/699/2/1473}, \href {https://ui.adsabs.harvard.edu/abs/2009ApJ...699.1473J} {699, 1473}

\bibitem[\protect\citeauthoryear{{Kempton} \& {Knutson}}{{Kempton} \& {Knutson}}{2024}]{Kempton2024}
{Kempton} E. M.~R.,  {Knutson} H.~A.,  2024, \mn@doi [Reviews in Mineralogy and Geochemistry] {10.2138/rmg.2024.90.12}, \href {https://ui.adsabs.harvard.edu/abs/2024RvMG...90..411K} {90, 411}

\bibitem[\protect\citeauthoryear{{Klein}, {Jura}, {Koester}, {Zuckerman}  \& {Melis}}{{Klein} et~al.}{2010}]{Klein2010}
{Klein} B.,  {Jura} M.,  {Koester} D.,  {Zuckerman} B.,   {Melis} C.,  2010, \mn@doi [\apj] {10.1088/0004-637X/709/2/950}, \href {https://ui.adsabs.harvard.edu/abs/2010ApJ...709..950K} {709, 950}

\bibitem[\protect\citeauthoryear{{Klein}, {Jura}, {Koester}  \& {Zuckerman}}{{Klein} et~al.}{2011}]{Klein2011}
{Klein} B.,  {Jura} M.,  {Koester} D.,   {Zuckerman} B.,  2011, \mn@doi [\apj] {10.1088/0004-637X/741/1/64}, \href {https://ui.adsabs.harvard.edu/abs/2011ApJ...741...64K} {741, 64}

\bibitem[\protect\citeauthoryear{{Klein}, {Doyle}, {Zuckerman}, {Dufour}, {Blouin}, {Melis}, {Weinberger}  \& {Young}}{{Klein} et~al.}{2021}]{Klein2021}
{Klein} B.~L.,  {Doyle} A.~E.,  {Zuckerman} B.,  {Dufour} P.,  {Blouin} S.,  {Melis} C.,  {Weinberger} A.~J.,   {Young} E.~D.,  2021, \mn@doi [\apj] {10.3847/1538-4357/abe40b}, \href {https://ui.adsabs.harvard.edu/abs/2021ApJ...914...61K} {914, 61}

\bibitem[\protect\citeauthoryear{{Koester}}{{Koester}}{2009}]{Koester2009}
{Koester} D.,  2009, \mn@doi [\aap] {10.1051/0004-6361/200811468}, \href {https://ui.adsabs.harvard.edu/abs/2009A&A...498..517K} {498, 517}

\bibitem[\protect\citeauthoryear{{Koester}}{{Koester}}{2010}]{Koester2010}
{Koester} D.,  2010, \memsai, \href {https://ui.adsabs.harvard.edu/abs/2010MmSAI..81..921K} {81, 921}

\bibitem[\protect\citeauthoryear{{Koester}, {Girven}, {G{\"a}nsicke}  \& {Dufour}}{{Koester} et~al.}{2011}]{Koester2011}
{Koester} D.,  {Girven} J.,  {G{\"a}nsicke} B.~T.,   {Dufour} P.,  2011, \mn@doi [\aap] {10.1051/0004-6361/201116816}, \href {https://ui.adsabs.harvard.edu/abs/2011A&A...530A.114K} {530, A114}

\bibitem[\protect\citeauthoryear{{Koester}, {G{\"a}nsicke}  \& {Farihi}}{{Koester} et~al.}{2014}]{Koester2014}
{Koester} D.,  {G{\"a}nsicke} B.~T.,   {Farihi} J.,  2014, \mn@doi [\aap] {10.1051/0004-6361/201423691}, \href {https://ui.adsabs.harvard.edu/abs/2014A&A...566A..34K} {566, A34}

\bibitem[\protect\citeauthoryear{{Kupka}, {Zaussinger}  \& {Montgomery}}{{Kupka} et~al.}{2018}]{Kupka2018}
{Kupka} F.,  {Zaussinger} F.,   {Montgomery} M.~H.,  2018, \mn@doi [\mnras] {10.1093/mnras/stx3119}, \href {https://ui.adsabs.harvard.edu/abs/2018MNRAS.474.4660K} {474, 4660}

\bibitem[\protect\citeauthoryear{{Lai} et~al.,}{{Lai} et~al.}{2021}]{Lai2021}
{Lai} S.,  et~al., 2021, \mn@doi [\apj] {10.3847/1538-4357/ac1354}, \href {https://ui.adsabs.harvard.edu/abs/2021ApJ...920..156L} {920, 156}

\bibitem[\protect\citeauthoryear{{Lodato}}{{Lodato}}{2008}]{Lodato2008}
{Lodato} G.,  2008, \mn@doi [\nar] {10.1016/j.newar.2008.04.002}, \href {https://ui.adsabs.harvard.edu/abs/2008NewAR..52...21L} {52, 21}

\bibitem[\protect\citeauthoryear{{Lodders}}{{Lodders}}{2019}]{Lodders2019}
{Lodders} K.,  2019, \mn@doi [arXiv e-prints] {10.48550/arXiv.1912.00844}, \href {https://ui.adsabs.harvard.edu/abs/2019arXiv191200844L} {p. arXiv:1912.00844}

\bibitem[\protect\citeauthoryear{{Madhusudhan}}{{Madhusudhan}}{2019}]{Madhusudhan2019}
{Madhusudhan} N.,  2019, \mn@doi [\araa] {10.1146/annurev-astro-081817-051846}, \href {https://ui.adsabs.harvard.edu/abs/2019ARA&A..57..617M} {57, 617}

\bibitem[\protect\citeauthoryear{{Malamud} \& {Perets}}{{Malamud} \& {Perets}}{2020}]{Malamud2020}
{Malamud} U.,  {Perets} H.~B.,  2020, \mn@doi [\mnras] {10.1093/mnras/staa142}, \href {https://ui.adsabs.harvard.edu/abs/2020MNRAS.492.5561M} {492, 5561}

\bibitem[\protect\citeauthoryear{{Martin} et~al.,}{{Martin} et~al.}{2005}]{GALEX2005}
{Martin} D.~C.,  et~al., 2005, \mn@doi [\apjl] {10.1086/426387}, \href {https://ui.adsabs.harvard.edu/abs/2005ApJ...619L...1M} {619, L1}

\bibitem[\protect\citeauthoryear{{Melis} \& {Dufour}}{{Melis} \& {Dufour}}{2017}]{Melis2017}
{Melis} C.,  {Dufour} P.,  2017, \mn@doi [\apj] {10.3847/1538-4357/834/1/1}, \href {https://ui.adsabs.harvard.edu/abs/2017ApJ...834....1M} {834, 1}

\bibitem[\protect\citeauthoryear{{Melis}, {Farihi}, {Dufour}, {Zuckerman}, {Burgasser}, {Bergeron}, {Bochanski}  \& {Simcoe}}{{Melis} et~al.}{2011}]{Melis2011}
{Melis} C.,  {Farihi} J.,  {Dufour} P.,  {Zuckerman} B.,  {Burgasser} A.~J.,  {Bergeron} P.,  {Bochanski} J.,   {Simcoe} R.,  2011, \mn@doi [\apj] {10.1088/0004-637X/732/2/90}, \href {https://ui.adsabs.harvard.edu/abs/2011ApJ...732...90M} {732, 90}

\bibitem[\protect\citeauthoryear{{Melis}, {Klein}, {Doyle}, {Weinberger}, {Zuckerman}  \& {Dufour}}{{Melis} et~al.}{2020}]{Melis2020}
{Melis} C.,  {Klein} B.,  {Doyle} A.~E.,  {Weinberger} A.,  {Zuckerman} B.,   {Dufour} P.,  2020, \mn@doi [\apj] {10.3847/1538-4357/abbdfa}, \href {https://ui.adsabs.harvard.edu/abs/2020ApJ...905...56M} {905, 56}

\bibitem[\protect\citeauthoryear{{Nittler}, {McCoy}, {Clark}, {Murphy}, {Trombka}  \& {Jarosewich}}{{Nittler} et~al.}{2004}]{Nittler2004}
{Nittler} L.~R.,  {McCoy} T.~J.,  {Clark} P.~E.,  {Murphy} M.~E.,  {Trombka} J.~I.,   {Jarosewich} E.,  2004, Antarctic Meteorite Research, \href {https://ui.adsabs.harvard.edu/abs/2004AMR....17..231N} {17, 231}

\bibitem[\protect\citeauthoryear{{Noack}, {Dorn}  \& {Baumeister}}{{Noack} et~al.}{2024}]{Noack2024}
{Noack} L.,  {Dorn} C.,   {Baumeister} P.,  2024, \mn@doi [arXiv e-prints] {10.48550/arXiv.2410.08055}, \href {https://ui.adsabs.harvard.edu/abs/2024arXiv241008055N} {p. arXiv:2410.08055}

\bibitem[\protect\citeauthoryear{{O'Brien} et~al.,}{{O'Brien} et~al.}{2023}]{OBrien2023}
{O'Brien} M.~W.,  et~al., 2023, \mn@doi [\mnras] {10.1093/mnras/stac3303}, \href {https://ui.adsabs.harvard.edu/abs/2023MNRAS.518.3055O} {518, 3055}

\bibitem[\protect\citeauthoryear{{O'Brien} et~al.,}{{O'Brien} et~al.}{2024}]{OBrien2024}
{O'Brien} M.~W.,  et~al., 2024, \mn@doi [\mnras] {10.1093/mnras/stad3773}, \href {https://ui.adsabs.harvard.edu/abs/2024MNRAS.527.8687O} {527, 8687}

\bibitem[\protect\citeauthoryear{{Paquette}, {Pelletier}, {Fontaine}  \& {Michaud}}{{Paquette} et~al.}{1986}]{Paquette1986}
{Paquette} C.,  {Pelletier} C.,  {Fontaine} G.,   {Michaud} G.,  1986, \mn@doi [\apjs] {10.1086/191112}, \href {https://ui.adsabs.harvard.edu/abs/1986ApJS...61..197P} {61, 197}

\bibitem[\protect\citeauthoryear{{Raddi}, {G{\"a}nsicke}, {Koester}, {Farihi}, {Hermes}, {Scaringi}, {Breedt}  \& {Girven}}{{Raddi} et~al.}{2015}]{Raddi2015}
{Raddi} R.,  {G{\"a}nsicke} B.~T.,  {Koester} D.,  {Farihi} J.,  {Hermes} J.~J.,  {Scaringi} S.,  {Breedt} E.,   {Girven} J.,  2015, \mn@doi [\mnras] {10.1093/mnras/stv701}, \href {https://ui.adsabs.harvard.edu/abs/2015MNRAS.450.2083R} {450, 2083}

\bibitem[\protect\citeauthoryear{{Robert} et~al.,}{{Robert} et~al.}{2024}]{Robert2024}
{Robert} A.,  et~al., 2024, \mn@doi [\mnras] {10.1093/mnras/stae1859}, \href {https://ui.adsabs.harvard.edu/abs/2024MNRAS.533.1756R} {533, 1756}

\bibitem[\protect\citeauthoryear{{Rocchetto}, {Farihi}, {G{\"a}nsicke}  \& {Bergfors}}{{Rocchetto} et~al.}{2015}]{Rocchetto2015}
{Rocchetto} M.,  {Farihi} J.,  {G{\"a}nsicke} B.~T.,   {Bergfors} C.,  2015, \mn@doi [\mnras] {10.1093/mnras/stv282}, \href {https://ui.adsabs.harvard.edu/abs/2015MNRAS.449..574R} {449, 574}

\bibitem[\protect\citeauthoryear{{Rogers} et~al.,}{{Rogers} et~al.}{2024a}]{Rogers2024}
{Rogers} L.~K.,  et~al., 2024a, \mn@doi [\mnras] {10.1093/mnras/stad3557}, \href {https://ui.adsabs.harvard.edu/abs/2024MNRAS.527.6038R} {527, 6038}

\bibitem[\protect\citeauthoryear{{Rogers} et~al.,}{{Rogers} et~al.}{2024b}]{Rogers2024b}
{Rogers} L.~K.,  et~al., 2024b, \mn@doi [\mnras] {10.1093/mnras/stae1520}, \href {https://ui.adsabs.harvard.edu/abs/2024MNRAS.532.3866R} {532, 3866}

\bibitem[\protect\citeauthoryear{{Rudnick} \& {Gao}}{{Rudnick} \& {Gao}}{2003}]{Rudnick2003}
{Rudnick} R.~L.,  {Gao} S.,  2003, \mn@doi [Treatise on Geochemistry] {10.1016/B0-08-043751-6/03016-4}, \href {https://ui.adsabs.harvard.edu/abs/2003TrGeo...3....1R} {3, 659}

\bibitem[\protect\citeauthoryear{{Schmidt}, {Keller}, {Francis}  \& {Bessell}}{{Schmidt} et~al.}{2005}]{SkyMAPPER2005}
{Schmidt} B.~P.,  {Keller} S.~C.,  {Francis} P.~J.,   {Bessell} M.~S.,  2005, in American Astronomical Society Meeting Abstracts \#206. p. 15.09

\bibitem[\protect\citeauthoryear{{Science Software Branch at STScI}}{{Science Software Branch at STScI}}{2012}]{STScI2012}
{Science Software Branch at STScI} 2012, {PyRAF: Python alternative for IRAF}, Astrophysics Source Code Library, record ascl:1207.011

\bibitem[\protect\citeauthoryear{{Skrutskie} et~al.,}{{Skrutskie} et~al.}{2006}]{2MASS2006}
{Skrutskie} M.~F.,  et~al., 2006, \mn@doi [\aj] {10.1086/498708}, \href {https://ui.adsabs.harvard.edu/abs/2006AJ....131.1163S} {131, 1163}

\bibitem[\protect\citeauthoryear{{Subasavage}, {Henry}, {Bergeron}, {Dufour}, {Hambly}  \& {Beaulieu}}{{Subasavage} et~al.}{2007}]{Subasavage2007}
{Subasavage} J.~P.,  {Henry} T.~J.,  {Bergeron} P.,  {Dufour} P.,  {Hambly} N.~C.,   {Beaulieu} T.~D.,  2007, \mn@doi [\aj] {10.1086/518739}, \href {https://ui.adsabs.harvard.edu/abs/2007AJ....134..252S} {134, 252}

\bibitem[\protect\citeauthoryear{{Swan}, {Farihi}, {Koester}, {Hollands}, {Parsons}, {Cauley}, {Redfield}  \& {G{\"a}nsicke}}{{Swan} et~al.}{2019}]{Swan2019}
{Swan} A.,  {Farihi} J.,  {Koester} D.,  {Hollands} M.,  {Parsons} S.,  {Cauley} P.~W.,  {Redfield} S.,   {G{\"a}nsicke} B.~T.,  2019, \mn@doi [\mnras] {10.1093/mnras/stz2337}, \href {https://ui.adsabs.harvard.edu/abs/2019MNRAS.490..202S} {490, 202}

\bibitem[\protect\citeauthoryear{{Swan}, {Farihi}, {Melis}, {Dufour}, {Desch}, {Koester}  \& {Guo}}{{Swan} et~al.}{2023}]{Swan2023}
{Swan} A.,  {Farihi} J.,  {Melis} C.,  {Dufour} P.,  {Desch} S.~J.,  {Koester} D.,   {Guo} J.,  2023, \mn@doi [\mnras] {10.1093/mnras/stad2867}, \href {https://ui.adsabs.harvard.edu/abs/2023MNRAS.526.3815S} {526, 3815}

\bibitem[\protect\citeauthoryear{{Trierweiler}, {Doyle}, {Melis}, {Walsh}  \& {Young}}{{Trierweiler} et~al.}{2022}]{Trierweiler2022}
{Trierweiler} I.~L.,  {Doyle} A.~E.,  {Melis} C.,  {Walsh} K.~J.,   {Young} E.~D.,  2022, \mn@doi [\apj] {10.3847/1538-4357/ac86d5}, \href {https://ui.adsabs.harvard.edu/abs/2022ApJ...936...30T} {936, 30}

\bibitem[\protect\citeauthoryear{{Trierweiler}, {Doyle}  \& {Young}}{{Trierweiler} et~al.}{2023}]{Trierweiler2023}
{Trierweiler} I.~L.,  {Doyle} A.~E.,   {Young} E.~D.,  2023, \mn@doi [PSJ] {10.3847/PSJ/acdef3}, \href {https://ui.adsabs.harvard.edu/abs/2023PSJ.....4..136T} {4, 136}

\bibitem[\protect\citeauthoryear{{Vanderbosch} et~al.,}{{Vanderbosch} et~al.}{2020}]{Vanderbosch2020}
{Vanderbosch} Z.,  et~al., 2020, \mn@doi [\apj] {10.3847/1538-4357/ab9649}, \href {https://ui.adsabs.harvard.edu/abs/2020ApJ...897..171V} {897, 171}

\bibitem[\protect\citeauthoryear{{Vanderburg} et~al.,}{{Vanderburg} et~al.}{2015}]{Vanderburg2015}
{Vanderburg} A.,  et~al., 2015, \mn@doi [\nat] {10.1038/nature15527}, \href {https://ui.adsabs.harvard.edu/abs/2015Natur.526..546V} {526, 546}

\bibitem[\protect\citeauthoryear{{Vennes}, {Kawka}, {Klein}, {Zuckerman}, {Weinberger}  \& {Melis}}{{Vennes} et~al.}{2024}]{Vennes2024}
{Vennes} S.,  {Kawka} A.,  {Klein} B.~L.,  {Zuckerman} B.,  {Weinberger} A.~J.,   {Melis} C.,  2024, \mn@doi [\mnras] {10.1093/mnras/stad3370}, \href {https://ui.adsabs.harvard.edu/abs/2024MNRAS.527.3122V} {527, 3122}

\bibitem[\protect\citeauthoryear{{Veras} \& {Heng}}{{Veras} \& {Heng}}{2020}]{Veras2020}
{Veras} D.,  {Heng} K.,  2020, \mn@doi [\mnras] {10.1093/mnras/staa1632}, \href {https://ui.adsabs.harvard.edu/abs/2020MNRAS.496.2292V} {496, 2292}

\bibitem[\protect\citeauthoryear{{Veras}, {Leinhardt}, {Bonsor}  \& {G{\"a}nsicke}}{{Veras} et~al.}{2014}]{Veras2014}
{Veras} D.,  {Leinhardt} Z.~M.,  {Bonsor} A.,   {G{\"a}nsicke} B.~T.,  2014, \mn@doi [\mnras] {10.1093/mnras/stu1871}, \href {https://ui.adsabs.harvard.edu/abs/2014MNRAS.445.2244V} {445, 2244}

\bibitem[\protect\citeauthoryear{Virtanen et~al.,}{Virtanen et~al.}{2020}]{2020SciPy-NMeth}
Virtanen P.,  et~al., 2020, \mn@doi [Nature Methods] {10.1038/s41592-019-0686-2}, \href {https://rdcu.be/b08Wh} {17, 261}

\bibitem[\protect\citeauthoryear{{Vogt} et~al.,}{{Vogt} et~al.}{1994}]{Vogt1994}
{Vogt} S.~S.,  et~al., 1994, in {Crawford} D.~L.,  {Craine} E.~R.,  eds,  Society of Photo-Optical Instrumentation Engineers (SPIE) Conference Series Vol. 2198, Instrumentation in Astronomy VIII. p.~362, \mn@doi{10.1117/12.176725}

\bibitem[\protect\citeauthoryear{{Von Hippel}, {Kuchner}, {Kilic}, {Mullally}  \& {Reach}}{{Von Hippel} et~al.}{2007}]{vonHippel2007}
{Von Hippel} T.,  {Kuchner} M.~J.,  {Kilic} M.,  {Mullally} F.,   {Reach} W.~T.,  2007, \mn@doi [\apj] {10.1086/518108}, \href {https://ui.adsabs.harvard.edu/abs/2007ApJ...662..544V} {662, 544}

\bibitem[\protect\citeauthoryear{{Williams}, {G{\"a}nsicke}, {Swan}, {O'Brien}, {Izquierdo}, {Cutolo}  \& {Cunningham}}{{Williams} et~al.}{2024}]{Williams2024}
{Williams} J.~T.,  {G{\"a}nsicke} B.~T.,  {Swan} A.,  {O'Brien} M.~W.,  {Izquierdo} P.,  {Cutolo} A.~M.,   {Cunningham} T.,  2024, \mn@doi [\aap] {10.1051/0004-6361/202450509}, \href {https://ui.adsabs.harvard.edu/abs/2024A&A...691A.352W} {691, A352}

\bibitem[\protect\citeauthoryear{{Wilson}, {Farihi}, {G{\"a}nsicke}  \& {Swan}}{{Wilson} et~al.}{2019}]{Wilson2019}
{Wilson} T.~G.,  {Farihi} J.,  {G{\"a}nsicke} B.~T.,   {Swan} A.,  2019, \mn@doi [\mnras] {10.1093/mnras/stz1050}, \href {https://ui.adsabs.harvard.edu/abs/2019MNRAS.487..133W} {487, 133}

\bibitem[\protect\citeauthoryear{{Wright} et~al.,}{{Wright} et~al.}{2010}]{WISE2010}
{Wright} E.~L.,  et~al., 2010, \mn@doi [\aj] {10.1088/0004-6256/140/6/1868}, \href {https://ui.adsabs.harvard.edu/abs/2010AJ....140.1868W} {140, 1868}

\bibitem[\protect\citeauthoryear{{Wyatt}, {Farihi}, {Pringle}  \& {Bonsor}}{{Wyatt} et~al.}{2014}]{Wyatt2014}
{Wyatt} M.~C.,  {Farihi} J.,  {Pringle} J.~E.,   {Bonsor} A.,  2014, \mn@doi [\mnras] {10.1093/mnras/stu183}, \href {https://ui.adsabs.harvard.edu/abs/2014MNRAS.439.3371W} {439, 3371}

\bibitem[\protect\citeauthoryear{{Xu}, {Jura}, {Dufour}  \& {Zuckerman}}{{Xu} et~al.}{2016}]{Xu2016}
{Xu} S.,  {Jura} M.,  {Dufour} P.,   {Zuckerman} B.,  2016, \mn@doi [\apjl] {10.3847/2041-8205/816/2/L22}, \href {https://ui.adsabs.harvard.edu/abs/2016ApJ...816L..22X} {816, L22}

\bibitem[\protect\citeauthoryear{{Xu}, {Zuckerman}, {Dufour}, {Young}, {Klein}  \& {Jura}}{{Xu} et~al.}{2017}]{Xu2017}
{Xu} S.,  {Zuckerman} B.,  {Dufour} P.,  {Young} E.~D.,  {Klein} B.,   {Jura} M.,  2017, \mn@doi [\apjl] {10.3847/2041-8213/836/1/L7}, \href {https://ui.adsabs.harvard.edu/abs/2017ApJ...836L...7X} {836, L7}

\bibitem[\protect\citeauthoryear{{Xu}, {Dufour}, {Klein}, {Melis}, {Monson}, {Zuckerman}, {Young}  \& {Jura}}{{Xu} et~al.}{2019}]{Xu2019}
{Xu} S.,  {Dufour} P.,  {Klein} B.,  {Melis} C.,  {Monson} N.~N.,  {Zuckerman} B.,  {Young} E.~D.,   {Jura} M.~A.,  2019, \mn@doi [\aj] {10.3847/1538-3881/ab4cee}, \href {https://ui.adsabs.harvard.edu/abs/2019AJ....158..242X} {158, 242}

\bibitem[\protect\citeauthoryear{{Xu}, {Lai}  \& {Dennihy}}{{Xu} et~al.}{2020}]{Xu2020}
{Xu} S.,  {Lai} S.,   {Dennihy} E.,  2020, \mn@doi [\apj] {10.3847/1538-4357/abb3fc}, \href {https://ui.adsabs.harvard.edu/abs/2020ApJ...902..127X} {902, 127}

\bibitem[\protect\citeauthoryear{{Zuckerman} \& {Becklin}}{{Zuckerman} \& {Becklin}}{1987}]{Zuckerman1987}
{Zuckerman} B.,  {Becklin} E.~E.,  1987, \mn@doi [\nat] {10.1038/330138a0}, \href {https://ui.adsabs.harvard.edu/abs/1987Natur.330..138Z} {330, 138}

\bibitem[\protect\citeauthoryear{{Zuckerman}, {Koester}, {Melis}, {Hansen}  \& {Jura}}{{Zuckerman} et~al.}{2007}]{Zuckerman2007}
{Zuckerman} B.,  {Koester} D.,  {Melis} C.,  {Hansen} B.~M.,   {Jura} M.,  2007, \mn@doi [\apj] {10.1086/522223}, \href {https://ui.adsabs.harvard.edu/abs/2007ApJ...671..872Z} {671, 872}

\bibitem[\protect\citeauthoryear{{Zuckerman}, {Koester}, {Dufour}, {Melis}, {Klein}  \& {Jura}}{{Zuckerman} et~al.}{2011}]{Zuckerman2011}
{Zuckerman} B.,  {Koester} D.,  {Dufour} P.,  {Melis} C.,  {Klein} B.,   {Jura} M.,  2011, \mn@doi [\apj] {10.1088/0004-637X/739/2/101}, \href {https://ui.adsabs.harvard.edu/abs/2011ApJ...739..101Z} {739, 101}

\makeatother
\end{thebibliography}







\bsp	
\label{lastpage}
\end{document}